# The GALAH Survey: A New Sample of Extremely Metal-poor Stars Using a Machine-learning Classification Algorithm

Arvind C. N. Hughes[1,2,3,4], Lee R. Spitler[1,2,3,5], Daniel B. Zucker[1,2,3], Thomas Nordlander[6,7], Jeffrey Simpson[7,8], Gary S. Da Costa[6,7], Yuan-Sen Ting[6,9], Chengyuan Li[10], Joss Bland-Hawthorn[3,11], Sven Buder[3,6], Andrew R. Casey[12,13], Gayandhi M. De Silva[2,5], Valentina D'Orazi[14], Ken C. Freeman[6], Michael R. Hayden[3,11], Janez Kos[15], Geraint F. Lewis[11], Jane Lin[3,6], Karin Lind[16], Sarah L. Martell[3,8], Katharine J. Schlesinger[6], Sanjib Sharma[3,11], and Tomaž Zwitter[15]
the GALAH Collaboration

[1] School of Mathematical and Physical Sciences, Macquarie University, Sydney, NSW 2109, Australia; arvind.hughes@students.mq.edu.au
[2] Research Centre in Astronomy, Astrophysics & Astrophotonics, Macquarie University, Sydney, NSW 2109, Australia
[3] Centre of Excellence for Astrophysics in Three Dimensions (ASTRO-3D), Australia
[4] Max Planck Institute for Astronomy, Heidelberg, Germany
[5] Australian Astronomical Optics, Faculty of Science and Engineering, Macquarie University, Macquarie Park, NSW 2113, Australia
[6] Research School of Astronomy and Astrophysics, Australian National University, Canberra, ACT 2611, Australia
[7] ARC Centre of Excellence for All Sky Astrophysics in 3 Dimensions (ASTRO 3D),Australia
[8] School of Physics, UNSW, Sydney, NSW 2052, Australia
[9] Research School of Computer Science, Australian National University, Acton, ACT 2601, Australia
[10] School of Physics and Astronomy, Sun Yat-sen University, Zhuhai, Guangdong, People's Republic of China
[11] Sydney Institute for Astronomy, School of Physics, A28, The University of Sydney, Sydney, NSW 2006, Australia
[12] Monash Centre for Astrophysics, Monash University, Clayton, VIC 3800, Australia
[13] School of Physics and Astronomy, Monash University, Clayton, VIC 3800, Australia
[14] Istituto Nazionale di Astrofisica, Osservatorio Astronomico di Padova, vicolo dell'Osservatorio 5, I-35122, Padova, Italy
[15] Faculty of Mathematics and Physics, University of Ljubljana, Jadranska 19, 1000 Ljubljana, Slovenia
[16] Department of Astronomy, Stockholm University, AlbaNova University Centre, SE-106 91 Stockholm, Sweden
Received 2022 January 18; revised 2022 March 10; accepted 2022 March 20; published 2022 May 3

## Abstract

Extremely metal-poor (EMP) stars provide a valuable probe of early chemical enrichment in the Milky Way. Here we leverage a large sample of ∼600,000 high-resolution stellar spectra from the GALAH survey plus a machine-learning algorithm to find 54 candidates with estimated [Fe/H] ⩽ −3.0, six of which have [Fe/H] ⩽ −3.5. Our sample includes ∼20% main-sequence EMP candidates, unusually high for EMP star surveys. We find the magnitude-limited metallicity distribution function of our sample is consistent with previous work that used more complex selection criteria. The method we present has significant potential for application to the next generation of massive stellar spectroscopic surveys, which will expand the available spectroscopic data well into the millions of stars.

*Unified Astronomy Thesaurus concepts:* Galactic archaeology (2178); Stellar abundances (1577); Stellar classification (1589); Classification (1907); CEMP stars (2105); Population II stars (1284); Population III stars (1285); Galaxy stellar content (621); High resolution spectroscopy (2096)

*Supporting material:* machine-readable table

## 1. Introduction

Extremely metal-poor (EMP) stars ([Fe/H] ⩽ −3.0) are interesting stellar objects, as they provide a window into the history of the early universe. The EMP stars that exist today formed in environments with much less chemical enrichment than is typically found in the interstellar medium today. As a result, they record the chemical yields produced by the first generations of stars after the Big Bang, and thereby provide crucial clues to the properties of early supernovae and their progenitors. Hence, EMP stars are essentially a log of some of the earliest events in the Galaxy's chemical evolution.

The significance of metal-poor stars has been reviewed extensively (e.g., Beers & Christlieb 2005; Frebel & Norris 2015).

However, to date very few EMP stars have been discovered, especially considering the fact that entire observational surveys have been dedicated to that aim. Querying the high-resolution SAGA database (Suda et al. 2008) we see that only ∼1000 stars with [Fe/H] < −3, ∼200 with [Fe/H] < −3.5, and ∼50 with [Fe/H] < −4 have been found. As noted above, EMP stars offer a unique window into the chemical enrichment of the Milky Way as it was forming, yet their relative rarity constrains our ability to probe those early times. Hence, expanding the known sample is of critical importance for creating a comprehensive picture of the processes dominating the life cycle of stars and the interstellar medium in the early universe.

With the development of highly multiplexed astronomical spectrographs, many current stellar surveys are producing spectroscopic data sets that are too large for traditional analysis (e.g., RAVE, Steinmetz et al. 2020; APOGEE, Ahumada et al. 2020; GALAH, Buder et al. 2021). The next generation of surveys, including WEAVE (Dalton et al. 2014), 4MOST (de Jong et al. 2019) and the Sloan Digital Sky Survey (SDSS) V's Milky Way Mapper (Kollmeier et al. 2017), will expand the available spectroscopic data well into the millions of stars. Thanks to recent improvements in computing and statistical methods, however, we are now able to develop more refined tools and processes to sift through these huge data sets in order to reliably identify rare and interesting science targets, such as EMP stars. This paper seeks to identify EMP stars in the GALAH spectroscopic survey, and develops a novel machine-learning approach that could be used to identify other scarce objects in large astronomical data sets.

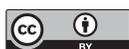

The machine-learning method adopted in this paper is t-distributed stochastic neighbor embedding (t-SNE; Maaten & Hinton 2008), a dimensionality-reduction technique. This method has been successfully applied to astronomical data for identification of substructure within a parameter space and the classification of stellar objects; in particular, t-SNE has been applied in the stellar and chemical abundance space, to identify membership in stellar clusters and streams (e.g., Anders et al. 2018; Kos et al. 2018). Matijevič et al. (2017) and Jofré et al. (2017) applied t-SNE to RAVE survey spectra to identify very metal-poor stars and stellar twins, respectively, and Traven et al. (2020) used t-SNE on GALAH spectra to find FGK-type binary stars. The application of t-SNE in these cases followed a top-down approach, i.e., running t-SNE on the data set in question and then exploring the resulting space, which can be inefficient in identifying objects of particular interest. In this paper, however, we show an alternative approach, following the process described in Hughes (2017) and similar to Hawkins et al. (2021), in which we flag objects we are interested in prior to running t-SNE, and then see where they appear on the projected t-SNE space. This works because any unclassified star falling near the flagged stars can be considered a potential candidate as it will have similar spectral features.

In this paper, data from the GALactic Archaeology with HERMES spectroscopic survey (GALAH) are analyzed to show how machine-learning methods, applied to spectra, can be used to identify EMP stars. The paper is organized as follows. The GALAH survey and the data are described in Section 2. Section 3 outlines the methodology used to find EMP stars, and the results and candidates from applying the methodology to the data are shown in Section 4. In Section 5 we discuss those results, and we summarize our conclusions in Section 6.

## 2. Data

The following section outlines the data sets used in this paper. Sections 2.1 and 2.2 briefly introduce the GALAH survey and discuss how the stars used in this analysis were selected. Section 2.3 describes how we label known EMP stars and other classified stars within GALAH. Lastly, Section 2.4 details how the synthetic templates that will be used in deriving stellar parameters are constructed.

### 2.1. Galactic Archaeology with the High Efficiency and Resolution Multi-Element Spectrograph

The GALAH survey is a high-resolution spectroscopic survey of the Milky Way which uses the High Efficiency and Resolution Multi-Element Spectrograph (HERMES) on the 3.9 m Anglo-Australian Telescope. By its conclusion, GALAH will obtain ~1,000,000 high-resolution spectra ($R \sim 28,000$) of stars at Galactic latitudes of $|b| > 10°$ and declinations $-80° < $ decl. $< +10°$, across the four discrete spectral arms of HERMES: 4713–4903 Å (blue channel), 5648–5873 Å (green channel), 6478–6737 Å (red channel), and 7585–7887 Å (IR channel). The spectrograph can typically achieve a signal-to-noise ratio (S/N) of ~100 per resolution element at magnitude $V \sim 14$ in the red arm during a 1 hr exposure (De Silva et al. 2015). By measuring radial velocities, stellar parameters, and abundances for as many as 30 elements, the goal of the GALAH survey is to produce a comprehensive view of the formation and chemodynamical evolution of the Milky Way.

This paper uses GALAH Data Release 3 (DR3; Buder et al. 2021), in which all observed stellar spectra were extracted as one-dimensional (1D) spectra, continuum normalized and radial velocity corrected to the barycentric reference frame. This data release includes spectra of ~600,000 unique stars. The GALAH data-reduction pipeline is described in Kos et al. (2017); for the data analysis pipeline, DR3 stellar parameters and abundances were estimated via the spectrum synthesis code Spectroscopy Made Easy (SME; Valenti & Piskunov 1996; Piskunov & Valenti 2017) using theoretical 1D hydrostatic models taken from the Marcs grid and 1D non-LTE grids, as described in Amarsi et al. (2020) for 11 elements (Li, C, O, Na, Mg, Al, Si, K, Ca, Mn, Fe, and Ba).

### 2.2. Sample Selection

We selected a subset of the ~600,000 DR3 stellar spectra tailored to the needs of our analysis. We considered spectra taken between 2013 November and 2018 February and limited ourselves to one spectrum per star, thus avoiding problems encountered with stacked spectra in DR3 (see Section 6.2 in Buder et al. 2021); in addition, we only used spectra that passed the reduction pipeline quality control (i.e., `red_flag==0`). We did not include poor-quality spectra with low signal-to-noise in the green channel (S/N < 35 per resolution element) and stars with GAIA $G_{BP} - G_{RP} < 0.6$ (Gaia Collaboration et al. 2018), as these typically represent hot stars that may appear to be EMP but are not.

### 2.3. Literature Stellar Labels

To be able to classify stellar spectra using semi-supervised machine learning, which is a combination of labeled and unlabelled data, we first have to assign a label to a proportion of the spectra. In our case the label is the stellar classification of the spectra.

A sample of stars with a known stellar classification was compiled by cross-matching the stellar classification labels defined by Traven et al. (2017) to the GALAH survey data using `s_object_ID`, a unique star identifier in DR3. The five stellar classes chosen based on SIMBAD are as follows:

1. Binary stars.
2. Cool metal-poor giants.
3. H$\alpha$/H$\beta$ emission stars.
4. Hot stars.
5. Stars with molecular absorption bands.

These stellar classifications were determined manually by Traven et al. (2017) after having run t-SNE in combination with Density-Based Spatial Clustering of Applications with Noise (SCAN or DBSCAN; Ester et al. 1996) on GALAH Data Release 1 (DR1; Martell et al. 2017). We therefore do not treat these labels as definite, as they only represent potential identifications. In addition, we defined two other categories of labeled objects: EMP stars (ExtMetalPoor), and one specific EMP star, the Keller star (Keller), which are described immediately below.

A sample of eight EMP stars was manually identified by cross-matching stars observed in GALAH with stars in SIMBAD determined to have [Fe/H] ≲ −3; this latter sample yielded a table with 538 unique entries. A 10″ positional cross-match of this table against the GALAH data resulted in a list of seven possible EMP stars based upon literature measurements compiled by SIMBAD. However, two of the stars failed to pass the GALAH quality checks because of poor spectrum normalization, and an additional star did not satisfy the metallicity requirement of [Fe/H] ∼ −3 as a detailed examination of literature measurements showed that it most likely has [Fe/H] ≈ −2. These spectra are given the label ExtMetalPoor.

The star SMSS J031300.36-670839.3 (Keller et al. 2014) was not a GALAH target, but was observed with HERMES on 2017 December 5 and then processed by the GALAH pipeline. The spectrum is almost featureless aside from hydrogen absorption, which is not surprising given its initial upper-limit estimate of [Fe/H] ≈ −7.1 (Keller et al. 2014), subsequently updated to [Fe/H] < −6.53 by Nordlander et al. (2017). This star is given the classification of Keller.

The sample of known EMP stars described above—the five ExtMetalPoor stars and the Keller star—are presented in Table 1, in which they are highlighted by a *.

The stars that do not have a stellar classification label after cross-matching by s_object_ID are given the classification unlabelled. These observations are combined with the labeled data set to define our full GALAH data set.

To summarize the GALAH data set used in this paper, after applying our sample-selection criteria for S/N and $G_{BP} - G_{RP}$ color, we produced a data set with 9058 labeled and 590,514 unlabelled stars. Thus the total number of stars analyzed in this work is 599,855.

### 2.4. Synthetic Templates

To determine the stellar parameters (see Section 3.2) of any potential EMP candidate we fit the observed spectra with synthetic spectral templates with known stellar parameters. Following Nordlander et al. (2019), the 6045 synthetic spectra templates were produced with Plez (2012) in 1D LTE using standard MARCS model atmospheres (Gustafsson et al. 2008). We used $v_{mic} = 1$ km s$^{-1}$ and plane-parallel model atmospheres for models with log $g > 3.5$, and $v_{mic} = 2$ km s$^{-1}$ and spherical geometry for models with log $g <= 3.5$. We fixed [α/H] = 0.4 and initially considered $T_{eff} = 4000$ K−8000 K in steps of 500K, log $g$ = 0.0–5.0 in steps of 0.5 and [Fe/H] = −7.0–0.0 in steps of 0.5 dex, and a limited range of carbon enhancements, [C/H] = 0.0, 0.5, 1.0. All synthetic spectra were broadened by $v_{broaden} = 10$ km s$^{-1}$ to represent the instrumental resolution, from an initial resolution of 1 km s$^{-1}$.

A finer grid was subsequently generated with $T_{eff} = 4000$ K−7500 K in steps of 100 K, log $g$ = 0.0–5.0 in steps of 0.25 dex, and varying step sizes in metallicity for different ranges of [Fe/H]:

1. −7.00 to −5.50 in steps of 0.5 dex;
2. −5.00 to −4.25 in steps of 0.25 dex;
3. −4.00 to −2.10 in steps of 0.1 dex; and
4. −2.00 to −1.00 in steps of 0.25 dex;

to give further detail on the range of [Fe/H] values that we could estimate. When applying the finer grid, we set the carbon abundance [C/Fe] = 0, as informed by our simulation analysis in the Appendix, and also because the wavelength ranges in GALAH cannot be used to meaningfully constrain the carbon abundance in EMP stars.

## 3. Method

Here we discuss a methodology that can be used to identify EMP stars within GALAH data, but also can potentially be adapted to find any stellar type within a given spectroscopic data set. The methodology is a hybrid of machine learning and a more traditional model-fitting approach. Section 3.1 outlines how to identify similar stars using a branch of machine learning known as dimensionality reduction, with a focus on targeting EMP stars. Once such candidate EMP stars have been found, Section 3.2 describes the estimation of their stellar parameters.

### 3.1. Identifying Similar Stars

Identification of EMP stars in spectroscopic surveys typically involves the fitting of select regions of an observed spectrum with a synthetic spectrum, employing a metric to define their similarity (usually $\chi^2$).

EMP stars have, however, a relatively featureless spectrum, where distinguishing the difference between a real spectral feature and the inherent noise is challenging. Similarly, the *synthetic* spectrum of a metal-poor star is close to featureless, but lacks the noise that is present in an observed spectrum. Hence, when applying a $\chi^2$-fitting method which entails comparing an observed spectrum with a synthetic spectrum, we expect stars that are not EMP to be identified as such (and vice versa) due to model systematics. Ideally, we would like a method that can (1) be independent of synthetic templates, (2) self-identify important features of a spectrum, and (3) categorise similar spectra.

To be able to create a method as described above is challenging for metal-poor stars. If, however, we could visualise the similarity of objects within a data set and group them visually, then we could reduce the search space for finding objects of interest, prior to running a $\chi^2$-fitting method. Furthermore, if we had a sample of the objects we were trying to identify, we could flag these before running a visualisation technique and see which groups they are clustered in, and hypothesise that the surrounding group must contain similar observations. By approaching the search for similar objects this way, we remove the necessity for synthetic template comparison at the identification phase of the process, resulting in a purer candidate sample.

Dimensionality-reduction techniques, a branch of machine learning, are a standard way of extracting important features from large data sets. Dimensionality reduction is the process of representing a high-dimensional data set, $X = x_1, x_2, ..., x_N$, by a set $Y$ of vectors, $y_i$, in two or three dimensions, and then placing similar observations in close proximity in the new parameter space, $y_i$, while keeping dissimilar observations at larger distances. The resulting reduced parameter space, $Y$, can then be visualised to determine similar and dissimilar input data.

**Table 1**
Extremely Metal-poor Stars from the Literature found in GALAH DR3, Designated with the GALAH Identifier `s_object_ID`

| s_object_ID | Object Name | $T_{eff}^L$ | log $g^L$ | [Fe/H]$^L$ | $T_{eff}^{Est}$ | log $g^{Est}$ | [Fe/H]$^{Est}$ | $T_{eff}^{DR3}$ | log $g^{DR3}$ | [Fe/H]$^{DR3}$ |
|---|---|---|---|---|---|---|---|---|---|---|
| 140209005201151* | HD 122563 Kirby et al. (2010) | 4367 | 0.60 | −3.15 | 5000 | 0.50 | −2.80 | 4616 | 1.46 | −2.51 |
| 140307003101095 | 2MASS J13274506-4732201 Simpson et al. (2012) | 4661 | 1.50 | −2.70 | 5000 | 0.50 | −2.10 | 4616 | 1.36 | −1.87 |
| 140412001201388 | HE 1207-3108 Yong et al. (2013) | 5294 | 2.85 | −2.70 | 5300 | 0.50 | −3.10 | 5404 | 2.97 | −2.56 |
| 140810004701232 | UCAC4 157-208544 Placco et al. (2019) | 4651 | 1.24 | −2.52 | 5000 | 0.50 | −2.10 | 4539 | 1.46 | −1.87 |
| 150409002601337 | TYC 4934-700-1 Sakari et al. (2018) | 4614 | 1.03 | −2.52 | 5100 | 0.50 | −2.60 | 4687 | 1.45 | −2.25 |
| 150718004401358* | BPS CS 22892-0052 McWilliam et al. (1995) | 4760 | 1.30 | −3.10 | 5200 | 3.75 | −3.50 | 5657 | 2.33 | −2.19 |
| 151008003501121* | HE 0124-0119 Li et al. (2015) | 4330 | 0.10 | −3.57 | 4000 | 5.00 | −4.25 | 4367 | 1.63 | −3.38 |
| 160401003901201 | DENIS J133748.8-082617 Sakari et al. (2018) | 4265 | 0.25 | −2.62 | 4800 | 0.50 | −2.40 | 4289 | 0.73 | −2.44 |
| 160403004201044 | 2MASS J13273676-1710384 Placco et al. (2019) | 5223 | 1.67 | −2.55 | 5200 | 0.75 | −2.60 | 5127 | 2.12 | −2.17 |
| 160424004701042 | UCAC4 053-017641 Placco et al. (2019) | 4832 | 1.61 | −3.41 | 5000 | 0.50 | −2.90 | 4795 | 2.05 | −2.51 |
| 160519002601142 | UCAC4 226-057537 Placco et al. (2019) | 4619 | 1.07 | −2.54 | 4900 | 0.50 | −2.30 | 4526 | 1.40 | −2.05 |
| 160813003601164* | 2MASS J21260896-0316587 Hollek et al. (2011) | 4725 | 1.15 | −3.22 | 5100 | 0.50 | −3.10 | 5056 | 2.15 | −2.71 |
| 161009903801062 | UCAC4 464-129364 Mardini et al. (2019) | 4945 | 1.53 | −2.52 | 5000 | 0.50 | −2.70 | 4743 | 2.10 | −2.36 |
| 161104002301201 | 2MASS J22045836 + 0401520 Spite et al. (2018) | 4700 | 1.20 | −2.90 | 5000 | 0.50 | −2.80 | 4632 | 1.82 | −2.57 |
| 161118004701028 | SMSS J051008.62-372019.8 Jacobson et al. (2015) | 5170 | 2.40 | −3.20 | 5300 | 0.50 | −3.20 | 5342 | 3.31 | −2.68 |
| 170601003101219 | 2MASS J14175995-2415463 Schlaufman & Casey (2014) | 4914 | 1.45 | −2.40 | 5000 | 0.50 | −2.60 | 4724 | 1.47 | −2.21 |
| 170615004401258* | 2MASS J18082002-5104378 Meléndez et al. (2016) | 5440 | 3.00 | −4.07 | 5500 | 0.50 | −4.25 | 5741 | 3.48 | ⋯ |
| 170805005101110 | HE 0048-6408 Placco et al. (2014a) | 4378 | 0.15 | −3.75 | 4800 | 0.50 | −3.80 | 4221 | 1.18 | −3.83 |
| 170904000601186 | 2MASS J21303218-4616247 Masseron et al. (2010) | 4100 | −0.30 | −3.39 | 5000 | 0.50 | −3.10 | 3987 | 0.89 | −3.76 |
| 170906004601038 | HE 0105-6141 Barklem et al. (2005) | 5218 | 2.83 | −2.55 | 5300 | 0.75 | −2.60 | 5190 | 2.87 | −2.36 |
| 170906004601108 | BPS CS 22953-0003 Spite et al. (2018) | 5100 | 2.30 | −2.80 | 5100 | 0.50 | −3.10 | 5044 | 2.36 | −2.73 |
| 171001003401116 | HE 0433-1008 Beers et al. (2017) | 4708 | 1.31 | −2.62 | 4900 | 0.50 | −2.70 | 4423 | 1.54 | −2.77 |
| 171205002101255* | SMSS J031300.36-670839.3 [Keller]Nordlander et al. (2017) | 5150 | 2.20 | <−6.53 | 5000 | ⋯ | ⋯ | ⋯ | ⋯ | ⋯ |

**Note.** The different superscripts in the stellar parameters reflect the source of the parameters: "L" indicates values from the literature, "Est" shows the results of our parameter estimation method and "DR3" represents the output of the GALAH DR3 pipeline. The stars marked with an asterisk are the "known" EMP stars used in the application of the methodology outlined in this paper; the remainder were subsequently identified as a verification of the method.

The most common dimensionality-reduction techniques used in astronomy are principal component analysis (PCA) and multidimensional scaling. PCA has been used by Yip et al. (2004) to classify quasars using SDSS spectral data; Connolly & Szalay (1999) demonstrated that PCA can be used to build galaxy spectral energy distributions (SEDs) from data that might be noisy or incomplete; and Re Fiorentin et al. (2007) showed that PCA can be used to estimate stellar atmospheric parameters with SDSS/SEGUE spectra, and Ting et al. (2012) used PCA to explore the stellar chemical abundance space. The frequent use of PCA underscores the importance of dimensionality reduction in the area of classification.

A significant weakness in PCA, however, is that it is intrinsically linear. PCA does not consider the structure of the manifold; there may be data points that form a nonlinear manifold, which PCA will not be able to deconstruct. In addition, while dimensionality-reduction techniques have been used in astronomy before, they generally were applied to smaller data sets. The effective parameter space for the GALAH data is 4 channels × 4096 pixels × 65,536 flux levels × 600,000 stars $\simeq 6 \times 10^{14}$ values; with a data set of this magnitude, traditional techniques face a computational challenge.

### 3.1.1. t-Distributed Stochastic Neighbor Embedding

Like other dimensionality-reduction techniques, t-SNE (Maaten & Hinton 2008) can be used to visualise how similar points are within a data set. t-SNE assesses the similarity of features in the higher-dimensional space by using the Euclidean distance metric (alternative metrics may be applied). A similarity matrix of probabilities, representing the higher-dimensional space, is calculated by converting these Euclidean distances using a standard Normal distribution. The feature space is then reduced to two or three dimensions, and the process above is repeated for this lower-dimensional space; however, the t-distribution is used instead of a standard Normal distribution to construct the similarity matrix. To finally determine the lower-dimensional representation of distances within our data set, the Kulback–Leibler (KL) divergence between the two joint-probability distributions is minimized using gradient descent. We chose the Barnes-Hut gradient descent version of t-SNE, implemented in the R package Rtsne[17] by Krijthe (2015), as it substantially speeds up t-SNE and allows t-SNE to be applied to much larger data sets that would be computationally intractable with the original t-SNE algorithm. t-SNE, unlike techniques such as PCA, is able to produce more visually compelling clusters because t-SNE 's nonlinearity enables it to maintain the trade-off between local and global similarities between points. This makes t-SNE well suited to the purpose of finding and visualising the distribution of similar spectra in a large data set; refer to Traven et al. (2017) for a more detailed description of the t-SNE algorithm.

To illustrate the effectiveness of t-SNE at classifying stars using only their spectra, we consider our defined labeled data set of 9058 classified stars. For this application, only the spectral data for the labeled stars was passed in to t-SNE. The labels and additional stellar parameter information were not used. t-SNE 's input is a set of 9058 high-dimensional objects $x_i, ..., x_N$, where each object is described by 12,288 wavelength values (for this analysis we ignore the IR channel), representing a single star. The top panel of Figure 1 shows the t-SNE map colored by effective temperature ($[T_{\rm eff}]$). The bottom panel of Figure 1 is the same map colored by the classification labels in Traven et al. (2017). In both panels, the cluster of hot stars is clearly distinguishable by both temperature and label, highlighting that by applying t-SNE to only spectra, we can visualise sensitivity in both the stellar parameter and stellar classification spaces.

### 3.1.2. Determining Which Wavelength Regions to Fit

In searching for EMP stars it is important to understand the significant spectral features which are key indicators of extremely low metallicity. When dealing with low- and medium-resolution spectra, traditionally the IR calcium triplet or ultraviolet calcium H and K lines have been used as standard indicators of low metallicity. These lines are, however, outside of the GALAH wavelength range. Therefore, the first step in applying our method to the entire GALAH data set is to determine which metal lines within the GALAH wavelength range would be most useful for identifying EMP stars.

The optimal wavelength ranges were selected by determining the lower limit for [Fe/H] using the spectral templates. This was achieved by running a series of $\chi^2$-fitting simulations using different elemental line combinations. The simulation which resulted in the highest level of certainty for the lowest [Fe/H] was selected. The optimal restricted wavelength ranges used are the regions around H$\alpha$, H$\beta$, 4867-4872 Å, and 4887-4892 Å; the latter two ranges contain the strongest FeI lines in the blue channel (4875.88 Å, 4890.76 Å, and 4891.49 Å). Additionally, we found that the OI triplet (7771.94 Å–7775.39 Å) in the IR channel was a useful discriminant for removing hot stars that were contaminating the sample, and thus this range was also included. In Figure 2 we show that, using spectra in the GALAH wavelength range, we can say with reasonable confidence that a star has a metallicity as low as [Fe/H] ∼ −3.5 (or potentially below that value); see the Appendix for further details.

By applying a method like t-SNE the idea is to reduce any bias that may arise in choosing which wavelength ranges to analyze, as the technique may be able to better identify significant wavelength ranges not considered. To use the entire wavelength range, however, is (a) computationally unfeasible, and (b), given the relatively featureless nature of EMP star spectra, can introduce noise that may skew the final results. We were thus unable to avoid having to select which wavelength ranges to input into t-SNE.

## 3.2. Estimating Stellar Parameters for EMP Stars

To confirm the identification of any EMP candidate found using the t-SNE methodology outlined in Section 3.1, we require an estimation of their basic stellar parameters, $T_{\rm eff}$, $\log g$, and [Fe/H]. The GALAH DR3 pipeline provides measurements of these stellar parameters for many of our candidates; however, the DR3 pipeline is not tailored toward metal-poor stars with weak metal lines. We developed a simple iterative procedure to estimate each stellar parameter for a candidate EMP star, which is described below (see Section 4.2 for the application).

### 3.2.1. Effective Temperature, Surface Gravity, and Metallicity

To estimate $T_{\rm eff}$, $\log g$, and [Fe/H], we apply a $\chi^2$-minimization routine between the observed spectra and the synthetic templates defined in Section 2.4 using only the H$\alpha$ and H$\beta$ regions and a select few metallicity lines around 4870 Å and 4890 Å. We fit the lines simultaneously to account for the degeneracies that arise between the stellar parameters, and select the template corresponding to the minimum $\chi^2$.

---
[17] https://github.com/jkrijthe/Rtsne

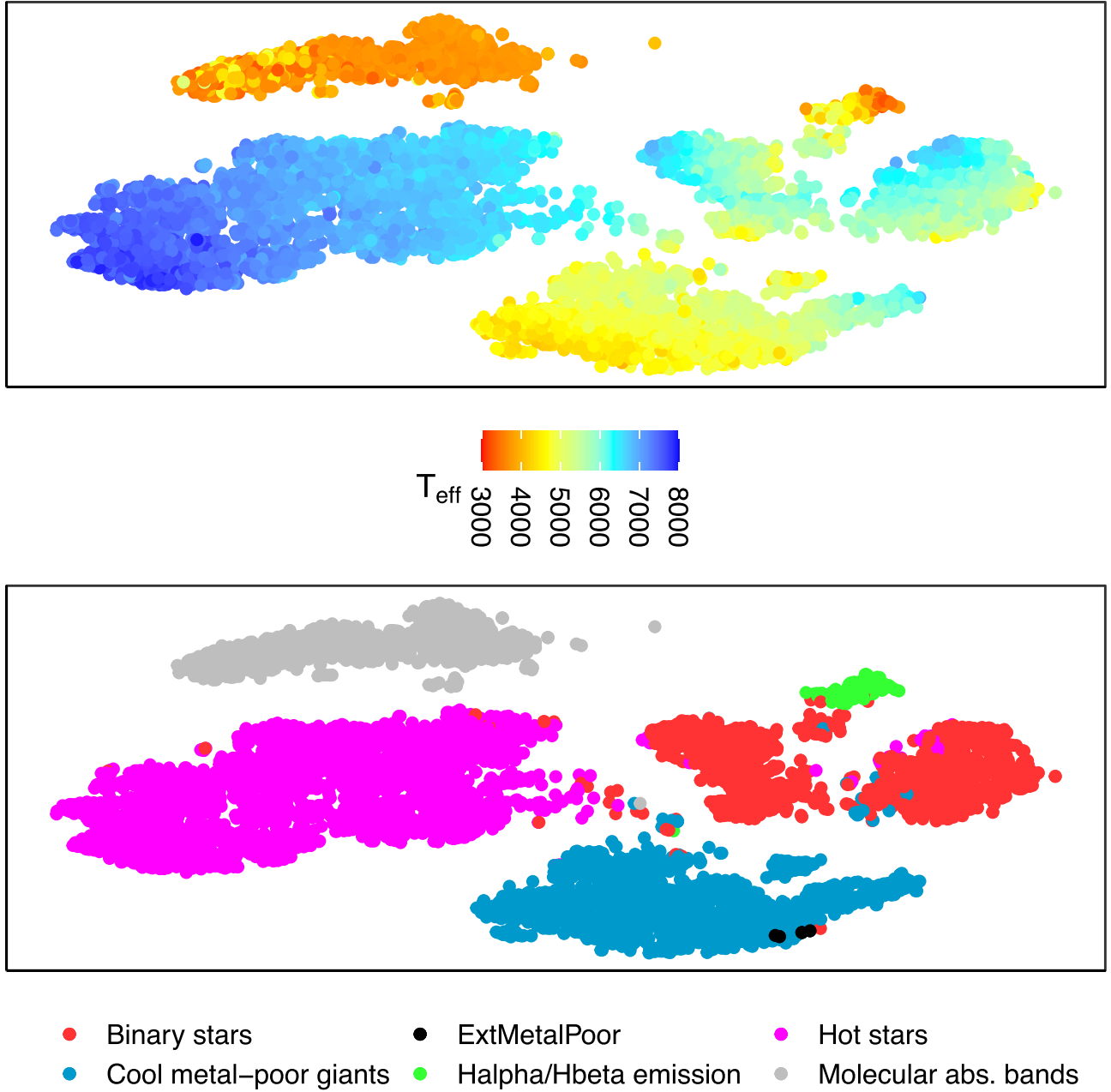

**Figure 1.** Illustrative t-SNE maps constructed only using the labeled portion our data set. The top panel is colored by effective temperature from GALAH DR3. In the bottom panel the t-SNE map is colored by stellar classification labels. Each point represents a star which has had its spectral information collapsed into two points in a two-dimensional parameter space produced by t-SNE; the axes are in arbitrary units reflecting only the dynamic range of the data in this space. Comparing the two panels shows that t-SNE is sensitive to effective temperature and the stellar classification. Note that neither spectroscopic temperature or classifications were included in the *input* to t-SNE.

The upper half of Figure 3 shows the fitting of the H$\beta$ region to three synthetic templates for a given star, with the optimal fit of $T_{\rm eff}$ equal to 5250, shown in red, and the best fit of log $g$ as 2.5. The bottom panels of Figure 3 show the two line regions considered in the fitting of metallicity, which suggest this observed star may have a metallicity in the range $-3.5 < $ [Fe/H] $ < -3.0$.

### 4. Results

The following section describes the results of applying the outlined methodology to the full GALAH data set inclusive of unclassified stars, as defined in Section 2.2. Section 4.1 applies our hybrid t-SNE methodology, and Sections 4.2 and 4.3 estimates the stellar parameters and applies some further analysis on the candidate sample, respectively.

#### 4.1. Applying the t-SNE Methodology

Applying the methodology to find EMP stars described in Section 3.1, we consider the entire GALAH sample, subsetted by the optimal wavelength regions as determined in Section 3.1.2 and with the EMP stars and the Keller star flagged. We will use the additional classification labels defined in Section 2.3 to flag other structures in the t-SNE plane.

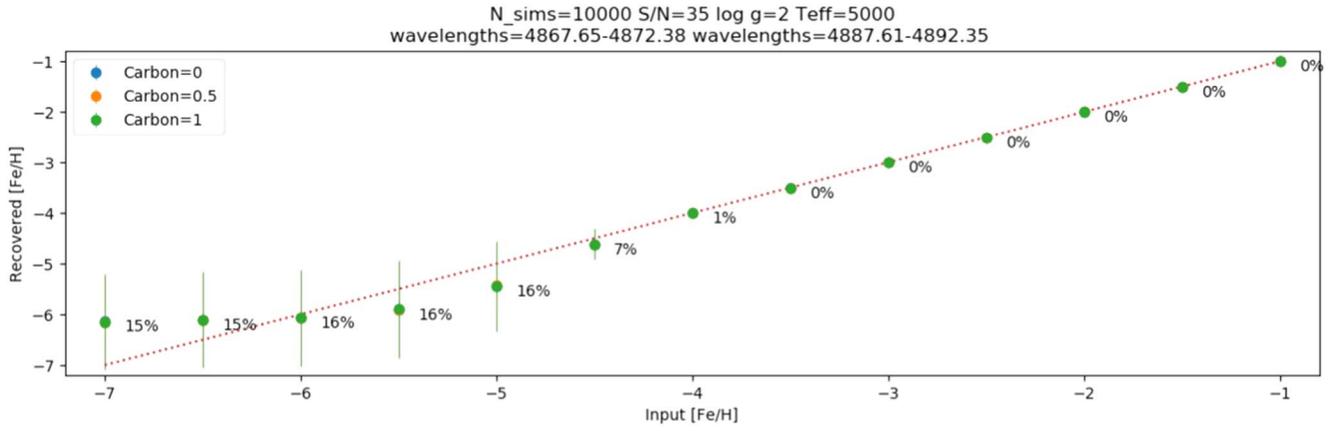

**Figure 2.** The output of a simulation fitting synthetic spectra with templates, showing that, given these stellar and observational parameters ($\log g = 2$, $T_{\rm eff} = 5000$ K, S/N = 35) we can reliably estimate metallicities down to [Fe/H] $\lesssim -3.5$ with GALAH data. Percentages indicate fractional uncertainty from scatter in the recovered metallicities rounded to the nearest percentage. Due to the relative insensitivity to carbon in the GALAH wavelength ranges, the blue and orange points are masked by the green points. See the Appendix for further details.

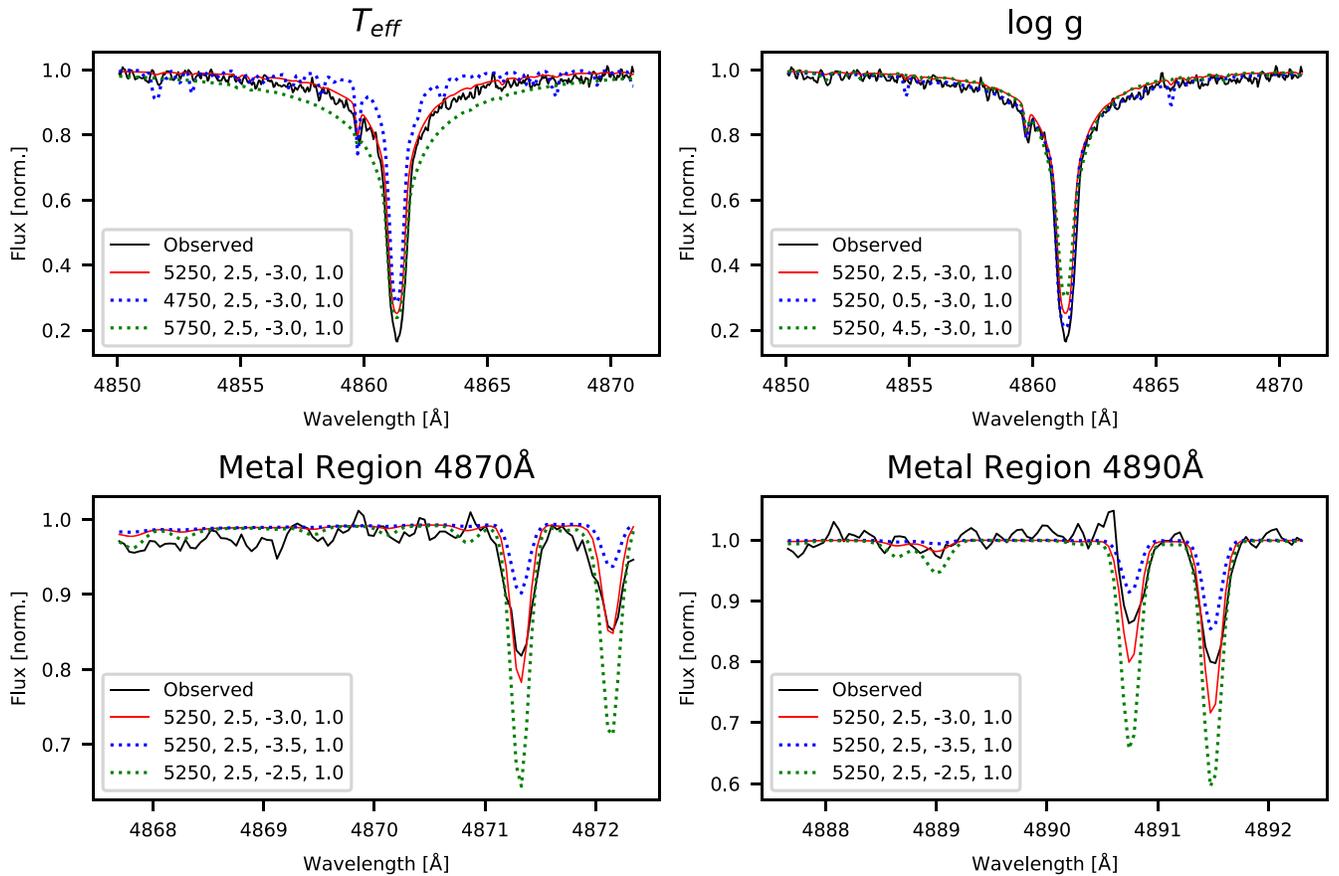

**Figure 3.** Fitting the observed H$\beta$ region for a given star with the wider synthetic template grid as a test of the fitting of the stellar parameters, $T_{\rm eff}$, $\log g$, and [Fe/H]. It is clear from the upper panels that this star is best fit by a $T_{\rm eff}$ of 5250 and a $\log g$ of 2.5. The bottom panels represent the line regions considered in fitting [Fe/H], and show that this star likely has a metallicity [Fe/H] between $-3.0$ and $-3.5$. The parameters of the synthetic templates as given in the panels are $T_{\rm eff}$, $\log g$, [Fe/H], and [C/Fe] (the assumed [C/Fe] may be ignored for these fits).

The t-SNE method was calculated with the perplexity set to 40, the number of iterations set to 2000, and the other hyperparameters (see Wattenberg et al. 2016) left to their default values. The processing was run on an Ubuntu server, with 344 GB of RAM and an Intel Xeon CPU E5-2695 v3 @ 2.30 GHz with 30 threads.

The resulting map is shown in Figure 4. A separate "island" containing all five known EMP stars and the Keller star is located in the top left of the map. We infer that the unlabelled stars surrounding the known Keller and EMP stars form a potential metal-poor cluster on the map. This cluster is then extracted and passed into our stellar parameter fitting routine described in Section 3.2, reducing the search space to fit stellar parameters of potential EMP stars from 600,000 to approximately 2500. A zoomed-in image of this cluster is shown in Figure 5.

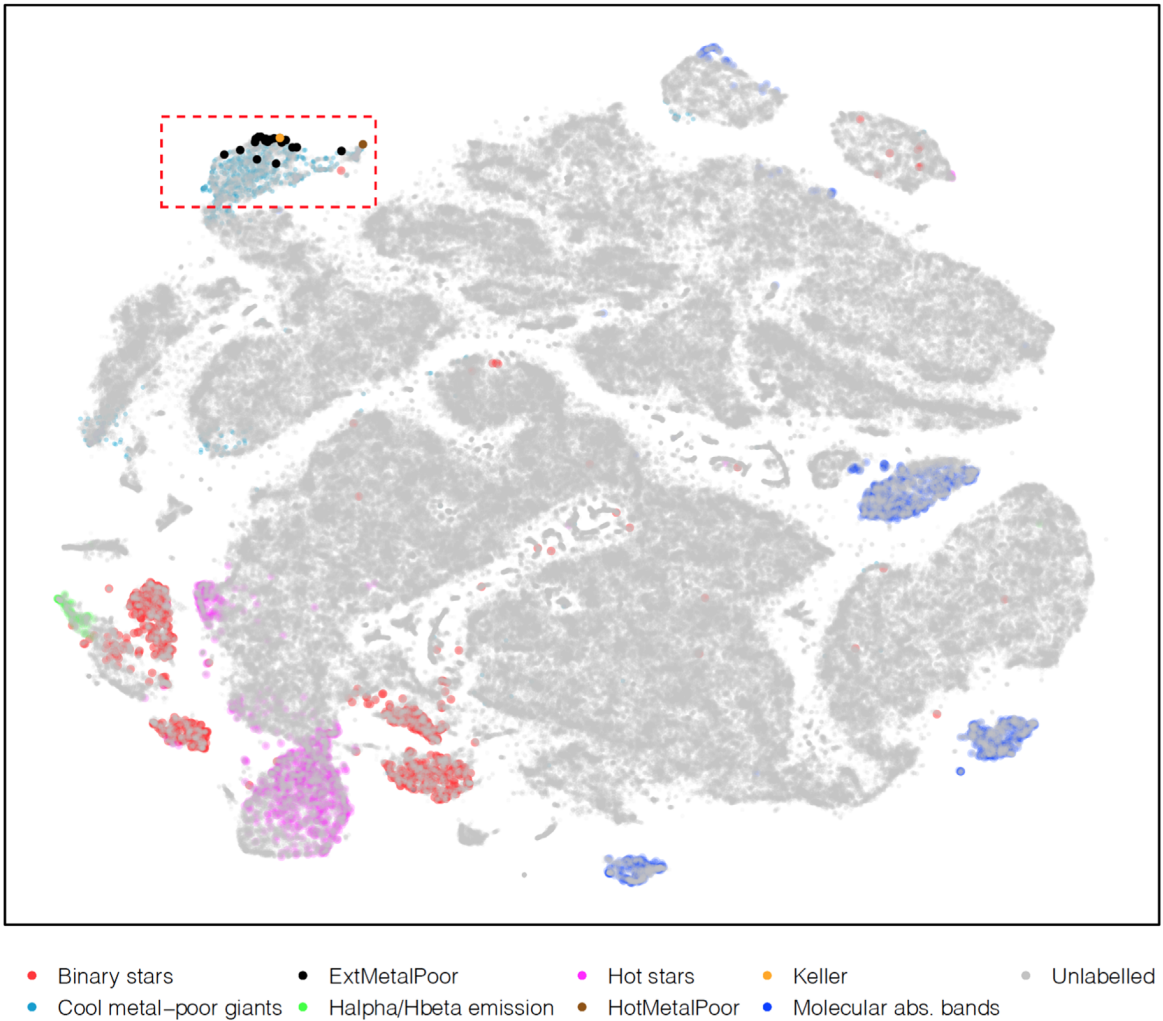

**Figure 4.** t-SNE map with the unknown (unlabelled) stars plotted in gray and the known extremely metal-poor (EMP) stars—corresponding to the stars shown in Table 1—overlaid in black, brown, and orange. A region containing all five (*) stars and the additional known metal-poor stars is located to the top left of the map. The dashed box represents the "island" selected for further analysis, with the EMP stars focused on the upper "coast" of the island.

### 4.2. Stellar Parameter Estimation for the EMP Stars Cluster

Taking the hypothesized metal-poor-only island, we estimate the stellar parameters for each star in the island using our simple stellar parameter fitting routine described in Section 3.2.

The estimated $T_{\rm eff}$ and $\log g$ values for our metal-poor island are shown in the upper two panels of Figure 6. Here we see a similar distribution of $T_{\rm eff}$ and $\log g$, with cooler giant-type stars on the left, going to hotter, higher $\log g$ stars to the right of the island. [Fe/H] is shown in the bottom panel of Figure 6 and displays a gradient of metallicity, higher to lower, from the bottom to the top edge of the island. The previously defined extremely metal-poor coast (as seen in Figure 5) is evident.

Before identifying and analyzing the EMP candidates in our cluster, we note that our stellar parameter fitting routine is relatively simple and is only used as a guide. This method was necessary, as we see a significant scatter with respect to DR3 pipeline-derived metallicities, as well as a systematic tendency toward higher measured metallicities in DR3. This may be attributed to the GALAH analysis pipeline being optimized for thin and thick disk stars, with typical metallicities $[{\rm Fe/H}] \geqslant -2$. Moreover, a comparison of our metallicity estimates for the stars with both the (admittedly heterogeneous) literature metallicities and GALAH DR3 metallicities, shown in Figure 7, suggests that our method yields reasonable estimates for [Fe/H]. Similar comparisons for our estimates of $T_{\rm eff}$ and $\log g$ (Figures 8 and 9, respectively) also show acceptable agreement with values from both the literature and GALAH DR3.

### 4.3. Candidates

Having used the template-fitting process described in Section 3.2 to estimate the stellar parameters of our cluster in Section 4.2, we find 380 stars that have $[{\rm Fe/H}] \leqslant -2.5$ and 135 stars with $[{\rm Fe/H}] \leqslant -2.7$. For the rest of the discussion, however, we only consider stars that have $[{\rm Fe/H}] \leqslant -3$, to satisfy the "extremely" metal-poor star designation. This results in 54 EMP candidates, six of which have an estimated $[{\rm Fe/H}] \leqslant -3.5$. We note that nine SIMBAD-sourced EMP stars from the literature in Table 1 are all contained in this t-SNE sample, and seven (two of which overlap with the literature) were identified as potential EMP stars by the GALAH DR3 pipeline, resulting in a net total of 40 potentially previously unidentified candidate EMP stars.

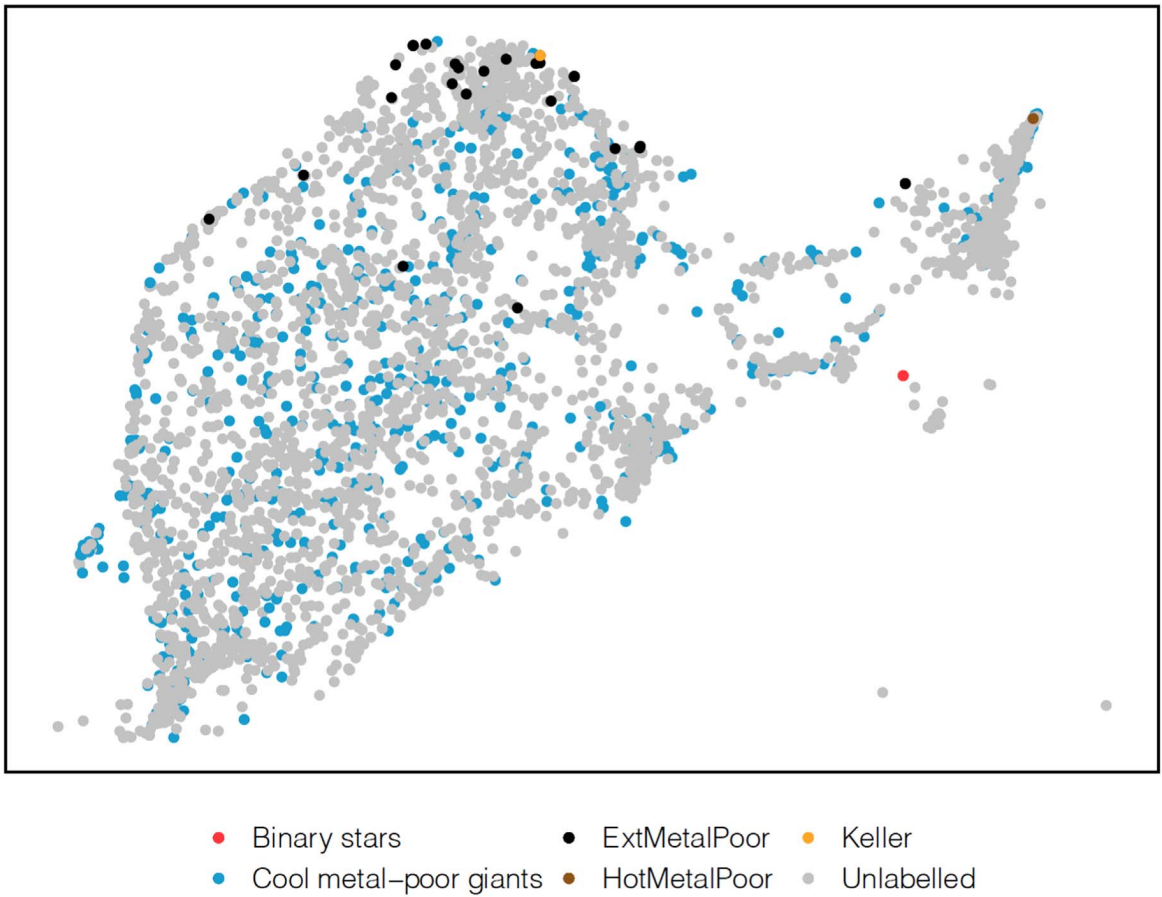

**Figure 5.** A zoomed-in view of the island highlighted by the red dashed box in Figure 4 with 2487 potential metal-poor stars. The known EMP stars lie on the upper extremely metal-poor "coast" of the island.

The spectra of the 54 candidate EMP stars are relatively featureless (with the exception of H$\alpha$ and H$\beta$) across the HERMES wavelength ranges; Table 2 shows the derived parameters for a few of our candidates. The spectrum of a representative EMP candidate from our selection is shown in Figure 10. This star has an [Fe/H] < −4.5, as determined by our stellar parameter routine.

As a first attempt to confirm that the candidates are likely EMP stars, we display their photometric properties in a parameter space that has successfully been used to select EMP stars. Figure 11 shows our sample cross-referenced with the SkyMapper photometric catalog (Onken et al. 2019). Here, $m_i$ represents a metallicity index, defined as $(v - g)_0 - 1.5(g - i)_0$, and $(g - i)_0$ as a proxy for $T_{\text{eff}}$. We show the EMP selection region in the figure from Da Costa et al. (2019), and find that most of our candidates are red giants and our sample fits within this region. This suggests, at least in terms of the broad metallicity-sensitive features targeted by the SkyMapper photometry, that our sample contains bona fide EMP stars. We note that Da Costa et al. (2019) found that 7% of the stars within the SkyMapper selection region ultimately proved to be EMP stars based on follow-up spectroscopy.

What about the candidates that fall *outside* of the selection box? We plot our candidates and the known literature stars on a color–magnitude diagram, using pho_g_mean_mag and the color $G_{\text{BP}} - G_{\text{RP}}$ from GAIA DR2 (Gaia Collaboration et al. 2018), and distances from GAIA (Bailer-Jones et al. 2018) in Figure 12. The majority of our candidates fall on the red giant branch along with some literature EMP stars, suggesting they are mostly red giants. Some of our candidates, however, are located near the main-sequence turn-off. A significant portion of our EMP candidates indeed show higher surface gravities, suggesting they are actually main-sequence turn-off stars.

## 5. Discussion

At a high level, given that we already had a large sample of high-resolution spectra, our candidate selection was relatively straightforward compared to previous EMP work: we have a magnitude-limited sample of stars and simply identified the population using iron and hydrogen absorption lines.[18] We note this is only possible because, even with the relatively limited wavelength coverage of GALAH spectra, that there is still sufficient sensitivity to spectral features indicative of EMP-like metallicities (see the Appendix).

One question which arises is how our sample compares to previous work on the metallicity distribution function (MDF) for EMP stars. The topic has been explored in a number of recent studies (e.g., Da Costa et al. 2019; Youakim et al. 2020; Yong et al. 2021), with Yong et al. (2021) finding a slope for the MDF of $\Delta(\log N)/\Delta[\text{Fe/H}] = 1.51$ dex per dex for $-4.0 <$ [Fe/H] $< -3.0$, with an apparent steep drop-off below $-4.0$ (below −4.0 it would appear virtually all stars are C-enhanced, with the [Fe/H] values likely varying stochastically depending on Population III supernova yields). The left panel of Figure 13 shows the MDF for our candidate sample and the right panel shows a log-scaled histogram with the gradient of 1.51 from Yong et al. (2021) overlaid (red dashed line). Here we can see that the "unbiased" nature of the current sample, which provides another way of investigating the form of the MDF, yields reasonably consistent results. However, given the MDF presented in Yong et al. (2021) and the current sample size (50 stars with [Fe/H] < −3.0), the probability of any of the current EMP candidates having [Fe/H] < −4.0 is not very high, as most will be closer to −3.0. Hence, a significantly larger EMP sample is required for probing the low-metallicity end of the stellar MDF; in this paper we have demonstrated that applying our approach to larger samples reaching fainter magnitudes is a key way to generate such an EMP sample.

---

[18] As noted previously, we also used an oxygen feature, but only as a discriminant to remove hot stars that were contaminating the sample.

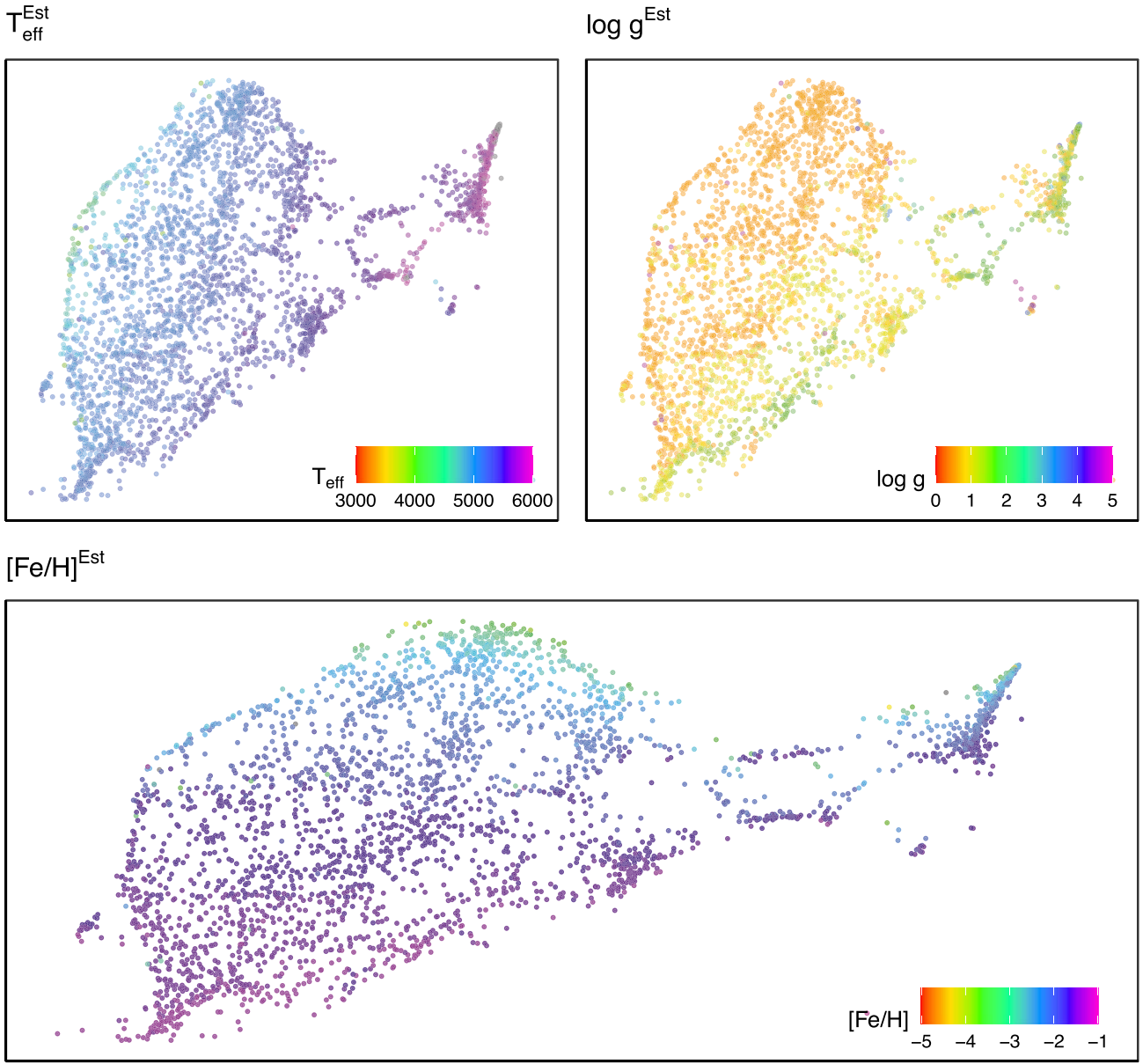

**Figure 6.** Three panels showing the selected island colored by each stellar parameter. The top-left panel shows our estimated $T_{\rm eff}$, having a distribution of temperature from cold to hot going left to right. Similarly, the top-right panel shows the estimated $\log g$ with a similar left to right distribution. The lower panel shows estimated [Fe/H] having a gradient of high to low metallicity from bottom to the top, with the top edge in agreement with the previously seen "extremely metal-poor coast."

Although the sample requires further spectroscopic observations to confirm our stellar parameter estimates, its relatively unbiased nature means there are a number of promising properties of the sample that suggest the method developed here has some advantages over other techniques for finding and understanding the EMP population.

First, as shown in Figures 11 and 12, we appear to have identified some main-sequence or main-sequence turn-off candidates. This is interesting because the sample of EMP stars from the literature observed serendipitously by GALAH (see, e.g., Table 1) consists of essentially all giant stars, reflecting the fact that previous work (e.g., Starkenburg et al. 2017) prioritised probing larger volumes in order to obtain large samples of relatively rare EMP stars. For this reason most surveys specifically targeted stars with giant-like properties, whose high luminosities allow them to be studied at greater distances.

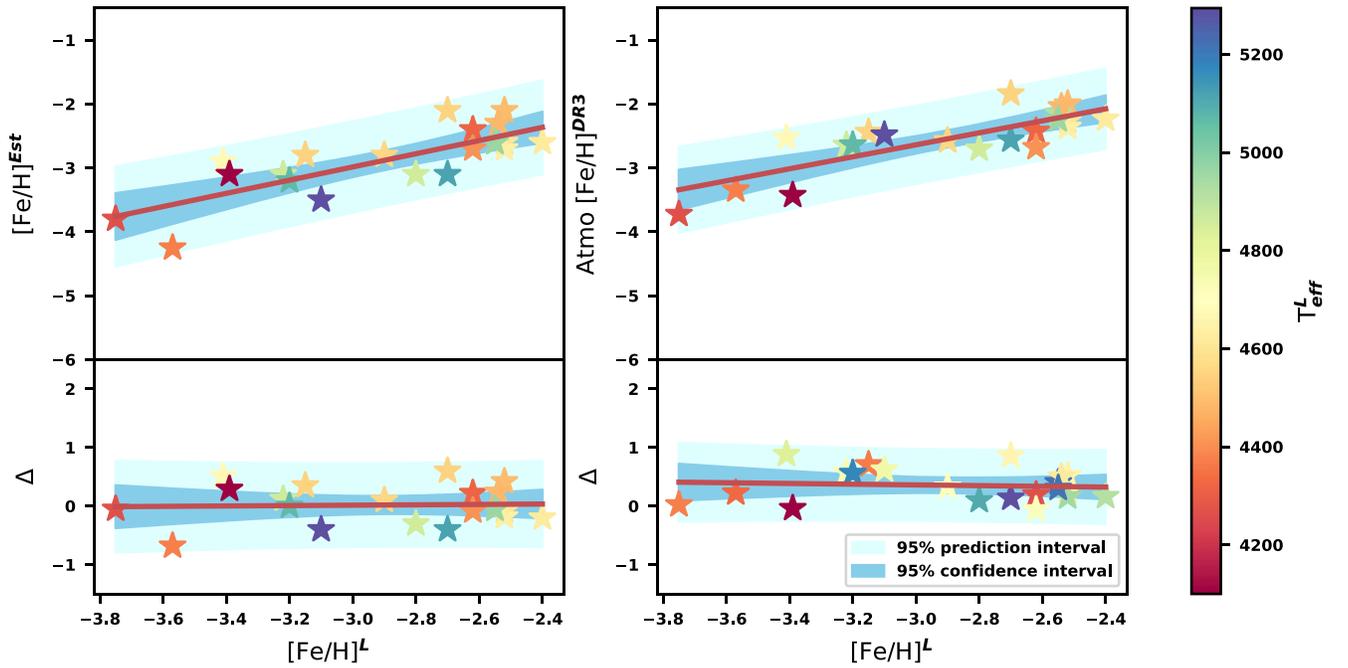

**Figure 7.** Two plots comparing our estimated [Fe/H] (left) and GALAH DR3 [Fe/H] (right) values with the literature. The top panels show the respective values while the bottom panels represent the difference between the method(s) and the literature. The red line shows a linear best fit to the data, with the prediction and confidence intervals as indicated by the shaded regions. Overall both methods have a 95% confidence band of approximately ±0.5, but the GALAH DR3 measured [Fe/H] values are higher on average than literature metallicities in this low-metallicity range.

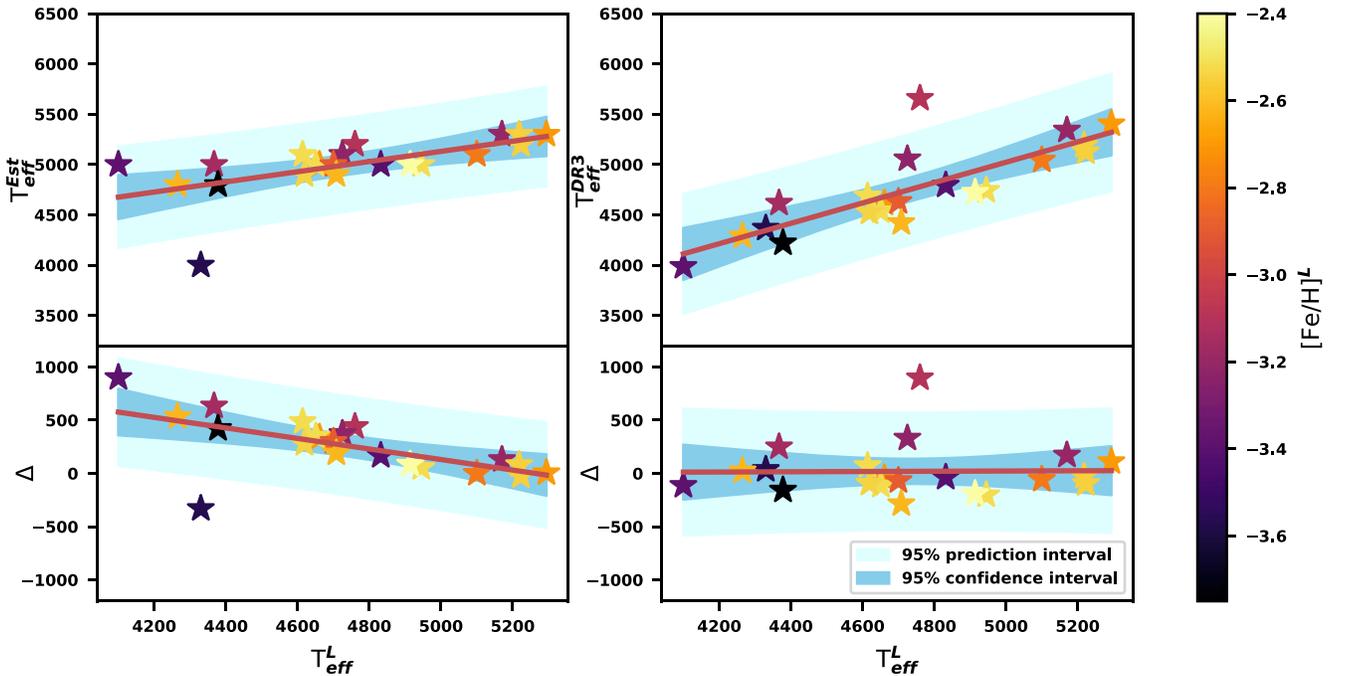

**Figure 8.** Two plots comparing our estimated $T_{\rm eff}$ (left) and GALAH DR3 $T_{\rm eff}$ (right) with the literature. The top panels show the respective values while the bottom panels represent the difference between the method(s) and the literature. The red line shows a linear best fit to the data, with the prediction and confidence intervals as indicated by the shaded regions. There is a trend in the errors of our estimation method, in that we have higher $T_{\rm eff}$ at the lower end, but overall have a similar error band to that of GALAH DR3.

If even one of our main-sequence EMP candidates turns out to be a bona fide main-sequence or main-sequence turn-off EMP star, this is an exciting opportunity to explore a less-studied population of EMP stars. The abundance patterns of main-sequence stars are comparatively easy to understand because they have not yet been affected by evolution in the post-main-sequence phase. The ages of these stars are also more accessible through comparison to isochrones, which is important for placing these EMP stars into the context of the formation and assembly of the Milky Way.

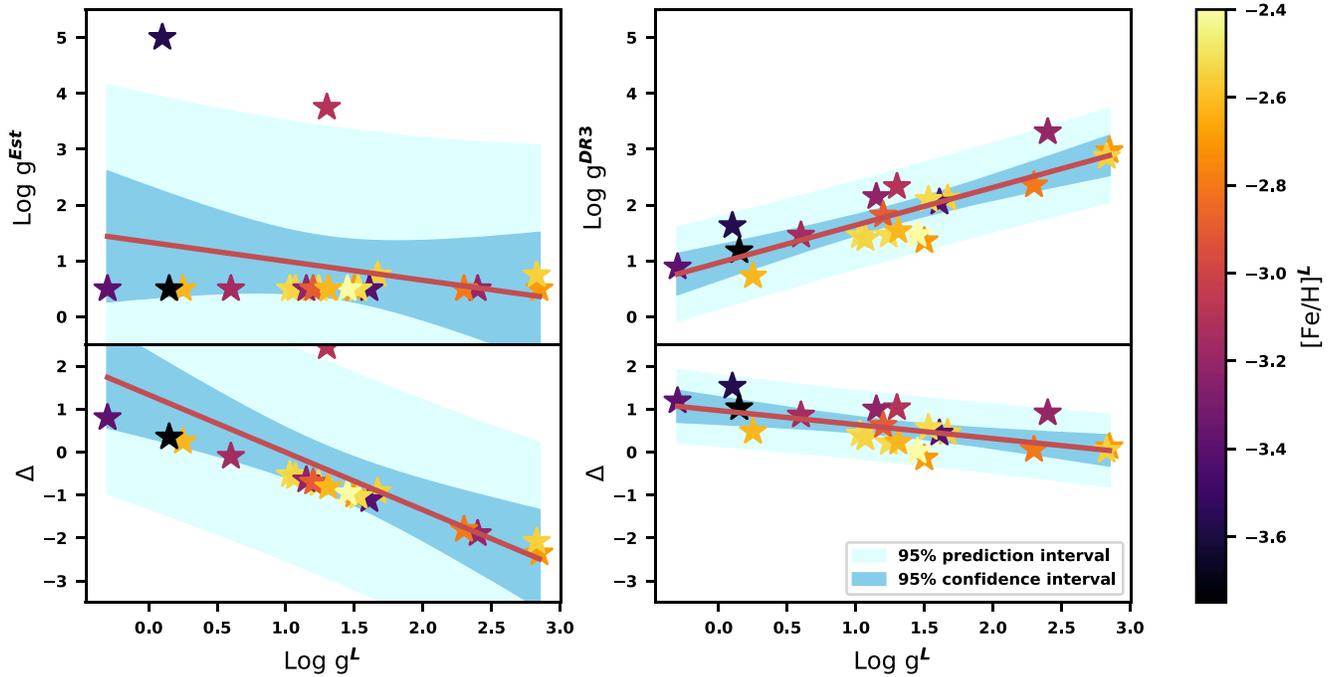

**Figure 9.** Two plots comparing our estimated log $g$ (left) and GALAH DR3 log $g$ (right) values with the literature. The top panels show the respective values while the bottom panels represent the difference between the method(s) and the literature. The red line shows a linear best fit to the data, with the prediction and confidence intervals as indicated by the shaded regions. Here you can clearly see that the GALAH DR3 estimates of log $g$ are a much better match, which is to be expected given the relative simplicity of our method.

**Table 2**
A Subset of EMP Candidates

| s_object_ID | R.A. | Decl. | $T_{eff}^{Est}$ | log $g^{Est}$ | [Fe/H]$^{Est}$ |
|---|---|---|---|---|---|
| 131123002501215 | 63.5677656 | −60.151311 | 5000 | 0.50 | −3.00 |
| 131217002301168 | 64.8334861 | −58.678350 | 5200 | 0.50 | −3.10 |
| 140312003501132 | 203.154833 | −38.009181 | 4900 | 0.50 | −3.30 |
| 140711001301222 | 242.630802 | −25.337563 | 5000 | 0.50 | −3.20 |
| 140808004701080 | 28.0619680 | −72.320519 | 5600 | 0.75 | −3.40 |
| 140809004901060 | 40.9968414 | −70.248597 | 4000 | 5.00 | −3.00 |
| … | … | … | … | … | … |

**Note**: The full candidate list is available electronically.

(This table is available in its entirety in machine-readable form.)

Another advantage of our method, which differs from other EMP selection methods, e.g., some combinations of photometric filters (Da Costa et al. 2019), is that carbon features did not affect our candidate selection. This means we have a relatively unbiased sample with respect to carbon abundance. Carbon-enhanced metal-poor stars (which have [C/Fe] > 0.7), become increasingly more frequent as [Fe/H] decreases (e.g., Placco et al. 2014b), and for [Fe/H]⩽−4.0, carbon-enhanced metal-poor stars dominate the known sample. Hence, this candidate sample presents an opportunity to explore the relative fraction of carbon-enhanced metal-poor stars as a function of [Fe/H] free of carbon-influenced selection bias. In fact, GALAH does not cover the wavelength ranges required to estimate carbon at extremely low metallicities, making follow-up observations of this magnitude-limited candidate sample essential for studying its carbon abundances.

The orbital information for our EMP candidates is captured in the vertical action and azimuthal action plot shown in Figure 14, similar to Figure 1 of Sestito et al. (2020) and Figure 5 of Cordoni et al. (2021). While our admittedly smaller set of candidates does not extend as high in vertical action as the Sestito et al. (2020) sample, we do see a significant near-"planar" component, biased toward prograde motion, in agreement with the results of both those authors and the SkyMapper-based study of Cordoni et al. (2021). Hence, while these kinematic data are not proof of the EMP nature of our candidates, they are consistent with the observed properties of confirmed EMP stars.

Finally, we note that the method presented here evolved from Hughes (2017), which employed t-SNE on GALAH spectra to classify them and identify several interesting classes of objects, including metal-poor stars. In that work, the fit used a relatively simple set of absorption features in the spectra. In the present work, we find a significant improvement in the quality of the candidates by assessing different line combinations in order to improve our metallicity sensitivity (see the Appendix). We furthermore include the GALAH IR fourth channel because it contains an oxygen feature—not previously considered—which served as a discriminant to reject spurious hot stars.

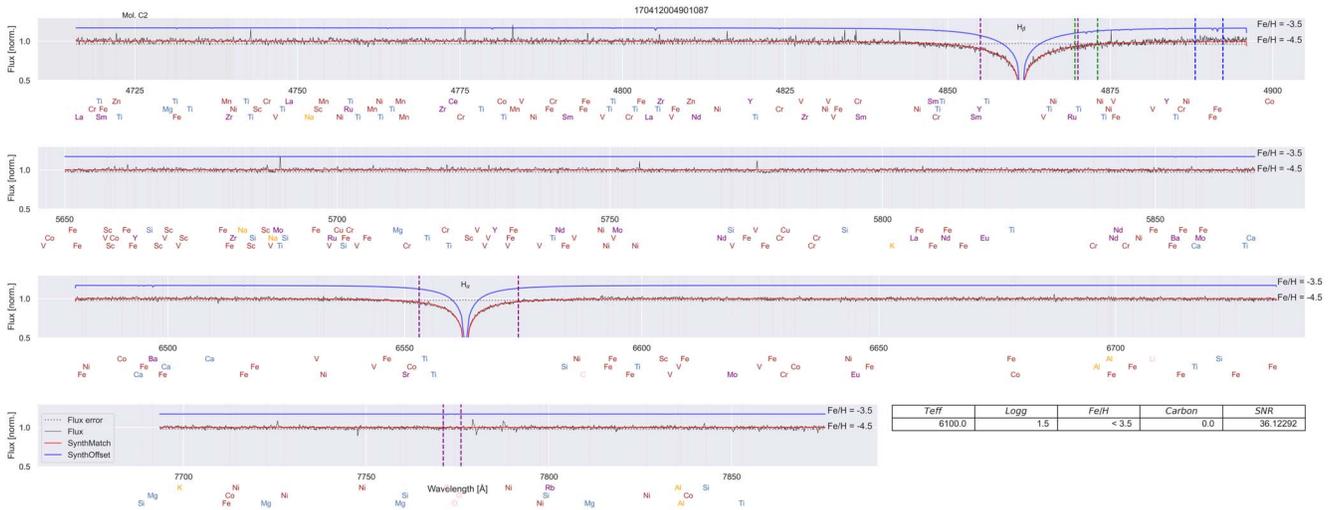

**Figure 10.** A candidate from the t-SNE EMP island identified in Figure 5 plotted over the four GALAH wavelength ranges, and overlaid with the best synthetic spectrum match in red, as well as a vertically shifted comparison spectrum offset by +1.0 in [Fe/H], shown in blue. The best-fit stellar parameters are listed in the bottom-right inset table and the vertical dashed lines represent the different wavelength regions used, as defined in Section 3.1.2.

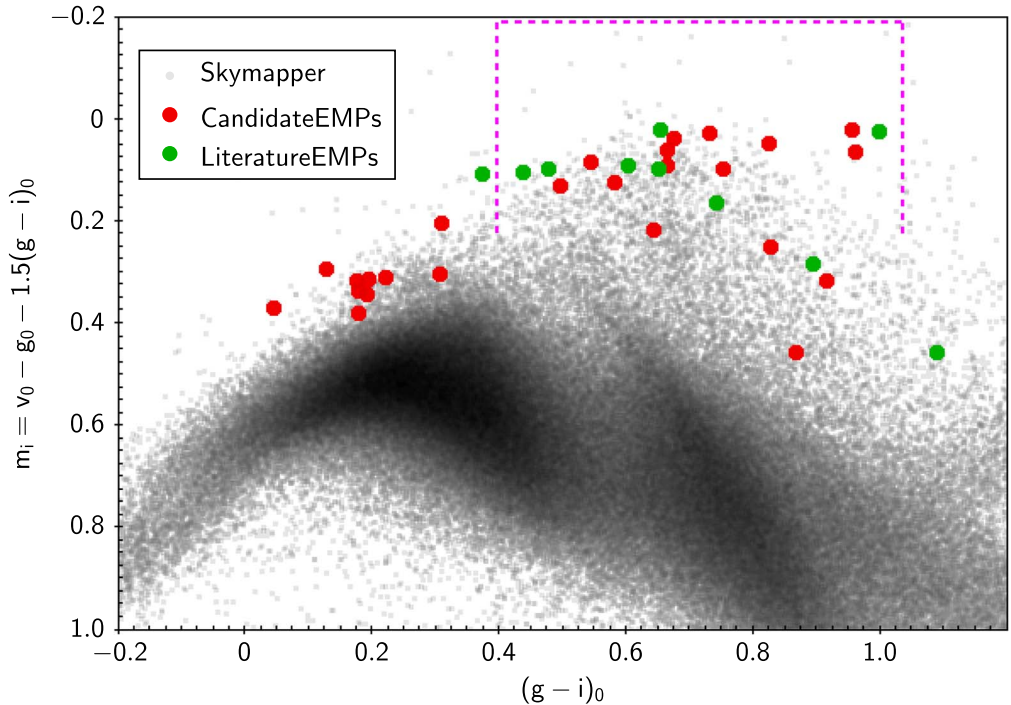

**Figure 11.** A SkyMapper metallicity-sensitive diagram, showing most of our candidates are likely red giants falling within the SkyMapper selection window (dashed magenta lines, from Da Costa et al. 2019). The compact grouping to the left represents candidate EMP main-sequence turn-off stars.

## 5.1. Advantages of a Machine-learning-based Approach Over More Traditional $\chi^2$-fitting Methods

As shown in Figure 5, our method uses t-SNE to isolate candidate EMP stars in a region with spectra similar to known EMP stars from the literature. By fitting synthetic spectra to the candidate EMP stars, we refine the selection of EMP candidates in the t-SNE space of EMP candidates to the top portion of the data shown in Figure 6. The clustering of known and candidate EMP stars in essentially a localised region in the entire t-SNE parameter space illustrates the potential power of our method. Nevertheless, a valid question is whether there are any improvements on our machine-learning-based method in terms of finding EMP stars over a simple $\chi^2$ fit to a wide range of synthetic spectral templates.

We tested the $\chi^2$ stellar parameter routine on the full GALAH data set, to potentially identify EMP stars that did not fall within the t-SNE EMP island identified in Figure 5. The results of this run are compared to the t-SNE run in Table 3. The purely $\chi^2$ method returned more potential candidates, but, upon visual inspection, 81% of those candidates were poorly fit, and some had strong absorption features, indicating that they are not good EMP candidates. The increased fraction of bad fits is likely because of model systematics—the minimum $\chi^2$ might not be representative of actual EMPs. Applying t-SNE before running a $\chi^2$-fitting routine minimizes this effect.

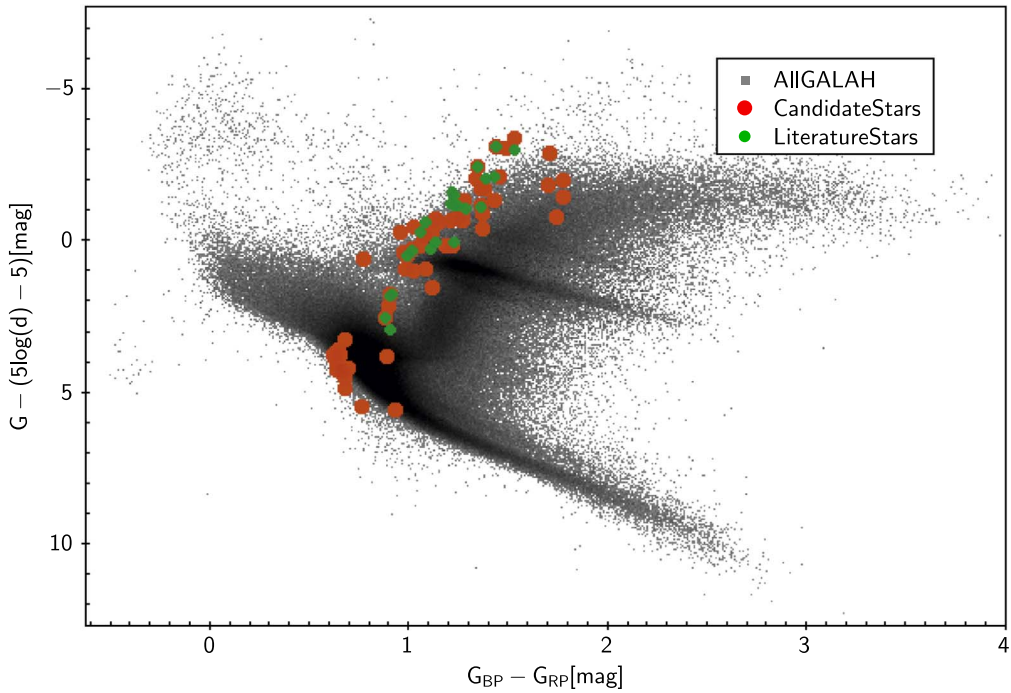

**Figure 12.** Color–magnitude diagram using magnitudes and distances from GAIA DR2 for the candidate EMP stars (yellow circles). The majority of the EMP candidates are red giants, while 20% appear to be consistent with main-sequence turn-off stars. Green circles are known EMP stars as defined in Table 1, including the most iron-poor star known, SMSS J031300.36670839.3 (Keller et al. 2014).

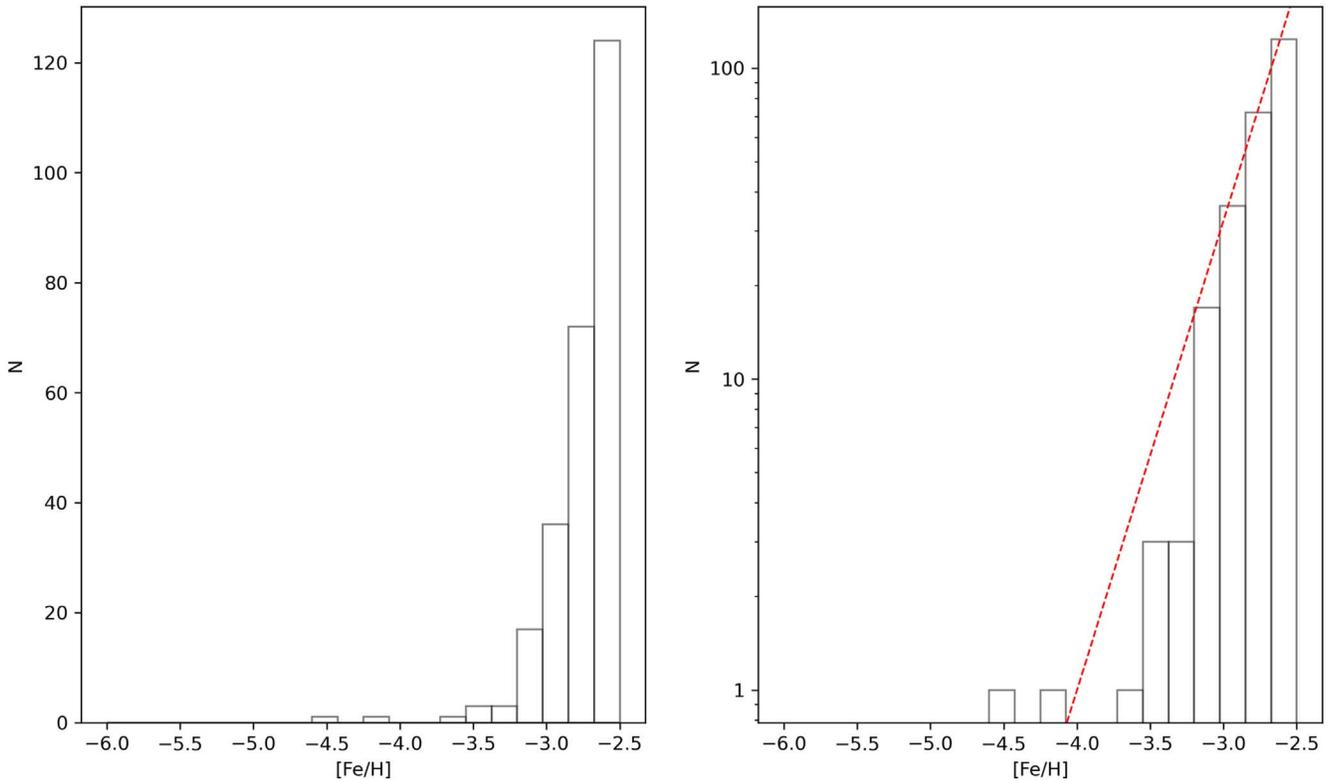

**Figure 13.** Metallicity distribution function for our candidate sample (left) and the log-scaled distribution function with the slope of 1.51 as determined in Yong et al. (2021) overlaid (red dashed line). For this histogram we only show candidates from the main GALAH survey, which is a magnitude-limited sample. We specifically exclude stars in GALAH DR3 targeted by other surveys (K2-HERMES, Wittenmyer et al. 2018; TESS-HERMES, Sharma et al. 2018; and GALAH-faint) because they incorporated fainter stars. The MDF follows a similar trend to that as seen in Yong et al. (2021), except for steeper fall-off at [Fe/H] < −3.3.

Moreover, we found that the $\chi^2$ method does not contain all the t-SNE EMP candidate sample: only 10 of our total sample of 60 (54 candidates and six spurious stars) are found. Finally,

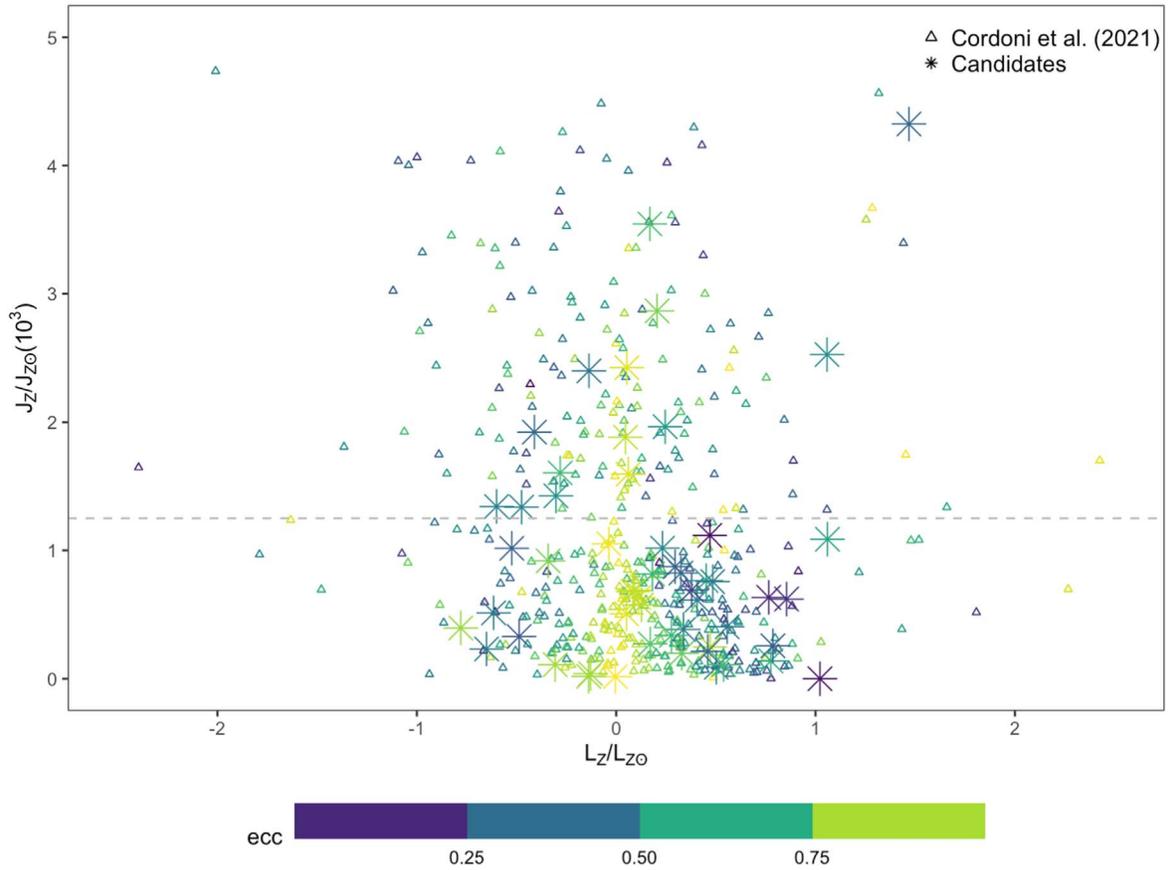

**Figure 14.** Vertical vs. azimuthal action components color-coded by eccentricity for our EMP candidates with metallicities [Fe/H] < −3 (star symbol), as well as literature values from Cordoni et al. (2021, triangle symbol). The action quantities are scaled by the solar values (i.e., $L_{z\odot} = 2009.92$ km s$^{-1}$ kpc, $J_{z\odot} = 0.35$ km s$^{-1}$ kpc). In this parameter space, we adopt the same horizontal dashed line at $J_z/J_{z\odot} = 1.25 \times 10^3$ as in Cordoni et al. (2021) to distinguish between planar and nonplanar orbits. The distribution of our candidates appears to be consistent with the observed orbital properties of confirmed EMP stars shown in Figure 1 of Sestito et al. (2020) and Figure 5 of Cordoni et al. (2021).

**Table 3**
Accuracy Percentages between our Method (i.e., t-SNE Classification, then a $\chi^2$ Fit to Models) and a Traditional $\chi^2$-fitting Technique for Finding EMP stars

| Method | EMP stars (%) | Extraneous sources (%) | Total EMP Candidates |
| --- | --- | --- | --- |
| $\chi^2$ only | 19 | 81 | 126 |
| t-SNE and $\chi^2$ | 90 | 10 | 60 |

**Note.** The percentage of EMP stars is the fraction of the total count that passed a visual inspection. Extraneous sources included both candidates with bad fits and those with strong absorption features, indicating that they are not likely to be EMP stars. The accuracy percentage of candidates that were found to be good EMP stellar candidates is higher using our method.

we also note that not all of the literature stars from Table 1 were recovered in the $\chi^2$ sample: only three of 23 are found.

## 6. Conclusions

We have demonstrated a methodology for finding EMP stars within a spectroscopic data set—in this case, spectra of ∼600,000 stars from the GALAH high-resolution survey—that is both computationally efficient and accurate, and may potentially be adapted to find other specific types of stars. Furthermore, we have shown that, using the GALAH wavelength ranges, we can derive metallicities down to [Fe/H] ∼ −3.5.

The candidate list we have identified is distinct from the results of many past surveys targeted specifically at EMP stars (e.g., Starkenburg et al. 2017; Da Costa et al. 2019). Given the nature of the GALAH data set—essentially a magnitude-limited sample of stellar spectra—our candidate list does not preferentially select giant stars (although, given their greater luminosity, giant stars probe a larger volume). This means we are sensitive to main-sequence and main-sequence turn-off stars, which are an interesting EMP population because, not having undergone dredge-ups, they are more likely to retain their original abundance patterns, and, in the case of main-sequence turn-off stars, they can potentially yield useful stellar ages. Moreover, the lack of strong carbon features in the GALAH wavelength windows means we are not biased against carbon-enhanced metal-poor stars—a significant fraction of EMP stars (Lee et al. 2013; Yong et al. 2013)—unlike some photometric-based EMP star surveys (see, e.g., Da Costa et al. 2019).

With regard to our methodology, we found a hybrid approach, i.e., preselection using t-SNE focused on specific wavelength regions, followed by parameter estimation via $\chi^2$ fitting, to be the most efficient way to identify candidate EMP stars. Simpler "brute-force" methods, for example applying t-SNE to the entire spectral range, or skipping machine-learning-based preselection and going straight to $\chi^2$ fitting to template spectra, proved to be both much more computationally intensive and much more likely to include extraneous spectra in their output. Although our method was tailored to GALAH spectra, we expect that similar techniques should be applicable to data sets from other ongoing and future large spectroscopic surveys, including WEAVE (Dalton et al. 2014) and 4MOST (de Jong et al. 2019).

While we have demonstrated that our metallicity estimates—along with those from the GALAH DR3 pipeline—are fairly reliable with regard to identifying EMP star candidates, follow-up observations, ideally covering additional regions of the optical spectrum more sensitive to low-metallicity measurements, are required to confirm these estimates, as well as to determine the abundances of carbon and other specific elements of interest (see, e.g., Beers & Christlieb 2005; Frebel & Norris 2015). To this end, we are engaged in a program of follow-up spectroscopy, with initial results expected shortly (G. Da Costa et al., in preparation).

We are grateful to the anonymous referee for their helpful comments and suggestions. Parts of this research were supported by the Australian Research Council Centre of Excellence for All Sky Astrophysics in 3 Dimensions (ASTRO 3D), through project number CE170100013. L.S. acknowledges support from Australian Research Council Discovery Project DP190102448. D.B.Z., J.S., and S.L.M. acknowledge support from Australian Research Council Discovery Project DP180101791. Y.S.T. acknowledges support from the Australian Research Council through DECRA Fellowship DE220101520. This work was based on data acquired at the Anglo-Australian Telescope. We acknowledge the traditional custodians of the land on which the AAT stands, the Gamilaraay people, and pay our respects to elders past and present.

*Facility:* AAT:HERMES .

*Software:* Astropy (Astropy Collaboration et al. 2013, 2018), Matplotlib (Hunter 2007, https://doi.org/10.1109/MCSE.2007.55), Rtsne (Krijthe 2015, https://github.com/jkrijthe/Rtsne).

## Appendix
## Deriving Metallicities for Metal-poor Candidates with GALAH Spectra

This section describes a set of simulation outputs that (1) illustrate the sensitivity of using only GALAH spectra to estimate metallicity at low metallicities ([Fe/H] $\sim -4.5$ is possible for cool giants), and (2) refine the line list we use for the low-metallicity fits (the three bluer channels are preferred slightly over a few strong Fe lines).

The simulations work by adding realistic noise to synthetic stellar templates with known parameters, and attempting to recover [Fe/H] using the same templates. We fix $[\alpha/H] = 0.4$ and consider a limited range of carbon enhancements ([C/H] = 0.0, 0.5, 1.0).

It is important to note that the following simulation results are only for fitting [Fe/H]. We assume that $\log g$ and $T_{\mathrm{eff}}$ are already well constrained (see Section 2.1 for how we do this with GALAH spectra) so we can limit the number of templates we fit to. The simulation results below did consider different carbon enhancements, in the sense that we explored whether carbon enhancement impacts the metallicity sensitivities. This means the current simulations were not meant to test our ability to constrain carbon abundance in an individual spectrum.

As shown in Section 4, the current sample of candidate EMP stars have a median S/N of 35 and two rough subpopulations: cool giants ($T_{\mathrm{eff}} = 5000$, $\log g \sim 2$) and hot main-sequence stars ($T_{\mathrm{eff}} = 6000$, $\log g \sim 4$).

To find the best regions of the GALAH spectra to fit for metallicity, we considered two different line lists as well as fitting to entire HERMES spectral channels. The first line list is from T. Nordlander and is highlighted shown in Figure 15. The second is a list of 57 metal-sensitive (mostly iron) lines compiled from features found in synthetic spectra and observed stars around [Fe/H] $\sim -3$ from K. Venn (private communication).

Figure 16 shows the output from one run of the simulation on the strongest features. Parameters for the input spectra are given in the title and the number of simulated stars for each input [Fe/H] was set to $N_{\mathrm{sims}} = 10{,}000$. The results show that for this spectral line that metallicity is well constrained until [Fe/H] ∼ −4.5. For lower metallicities the simulation indicates S/N = 35 spectral data over this line cannot distinguish between [Fe/H] ∼ −5 to −7 values.

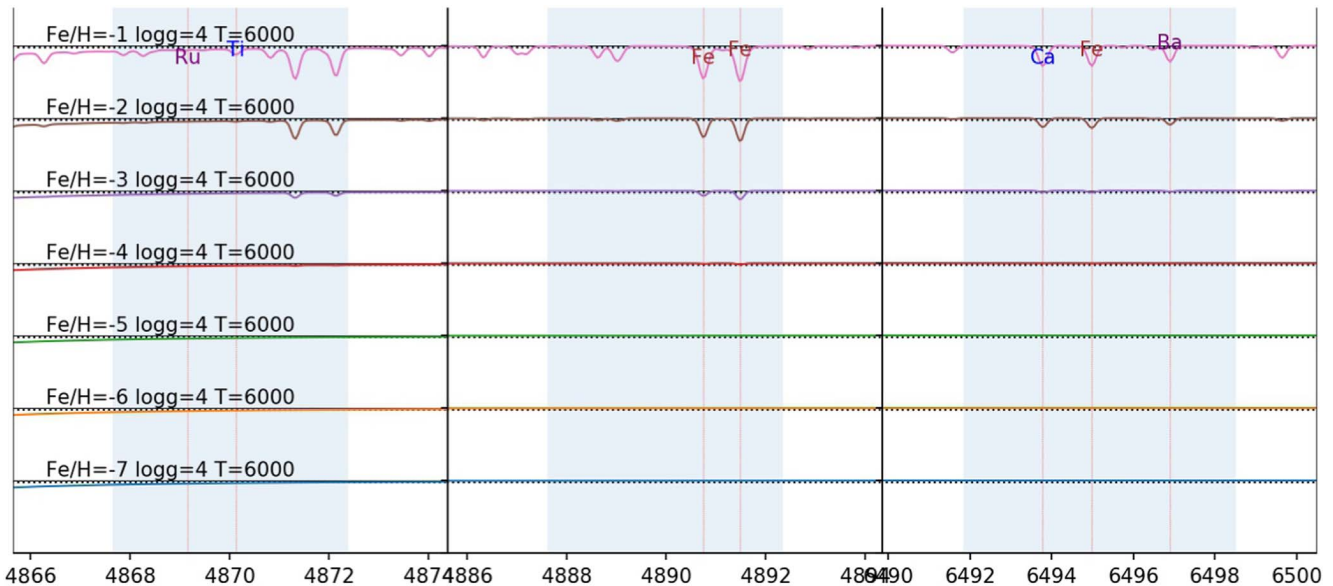

**Figure 15.** A plot of the synthetic spectra of a hot main-sequence star for a range of metallicities. The blue shaded areas are the wavelength regions with metal-sensitive absorption features.

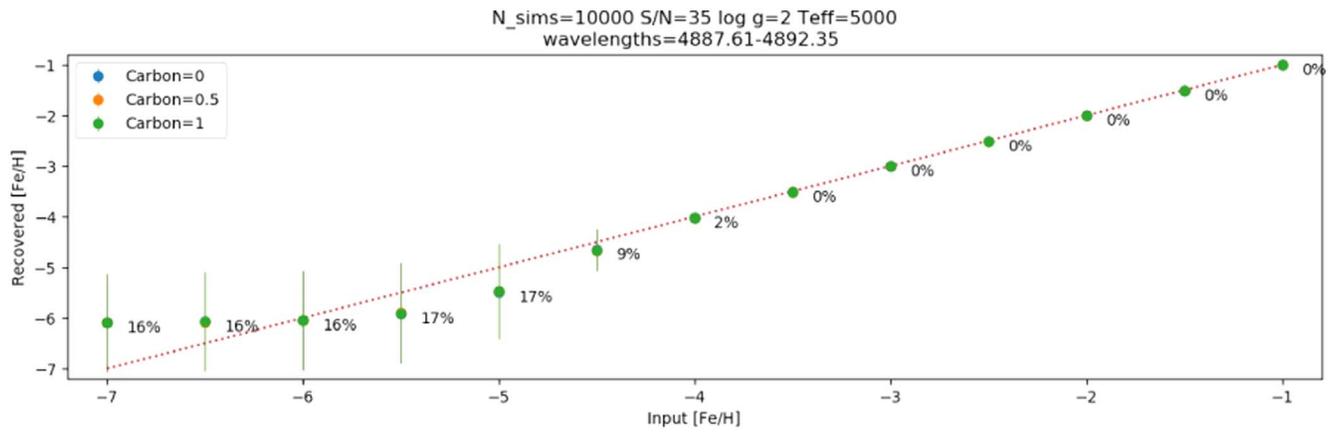

**Figure 16.** Simulation results for recovering input metallicities of synthetic cool, giant stellar spectra, with noise typical of the current sample. The points represent the recovered mean [Fe/H], with error bars reflecting the standard deviation in individual recovered [Fe/H] values. Percentages are the fractional uncertainty on the recovered metallicity. The data suggest we can constrain metallicity to within ∼ 8% down to [Fe/H] ∼ − 4.5 for cool giants and data with S/N = 35. For lower input metallicities, the simulation results indicate the spectra have essentially no constraint on metallicities [Fe/H] < − 4.5 at S/N = 35. The carbon abundances of the input spectra do not appear to impact the metallicity sensitivity, as can be seen by the carbon abundance of 1 (green) overlaying both the carbon abundance of 0 and 0.5 (blue and orange, respectively).

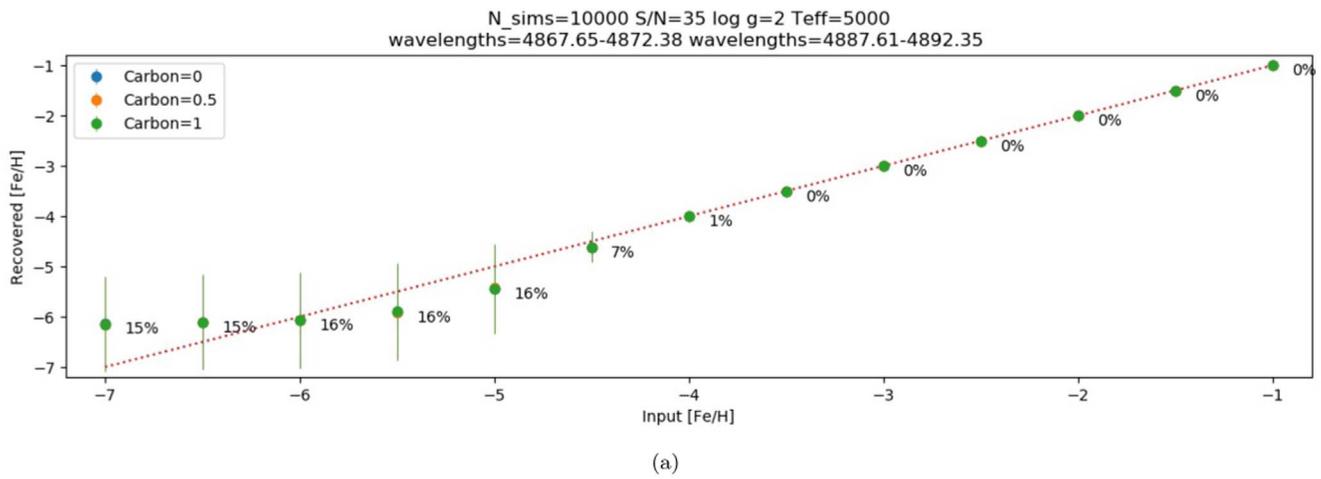

(a)

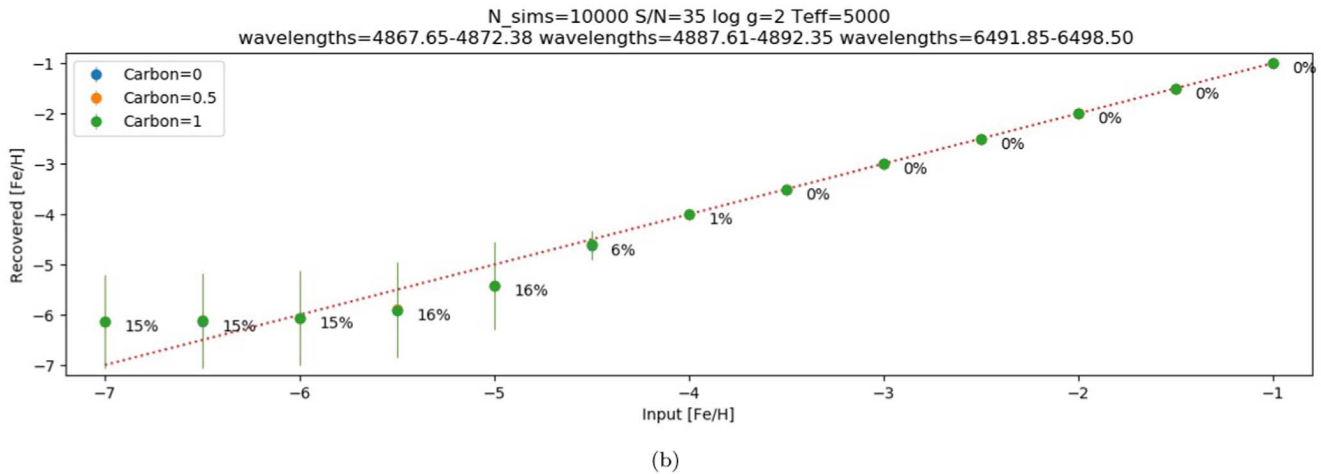

(b)

**Figure 17.** Same format as Figure 16, now with panels showing two different combinations of spectral lines shown in Figure 15, as indicated in the panel's subtitle. While the joint constraint of the spectral regions with the strongest features yields a better [Fe/H] constraint compared to a single line (see Figure 16), the third spectral region does not improve the fits for this stellar type and assumed S/N, likely because its features are weaker.

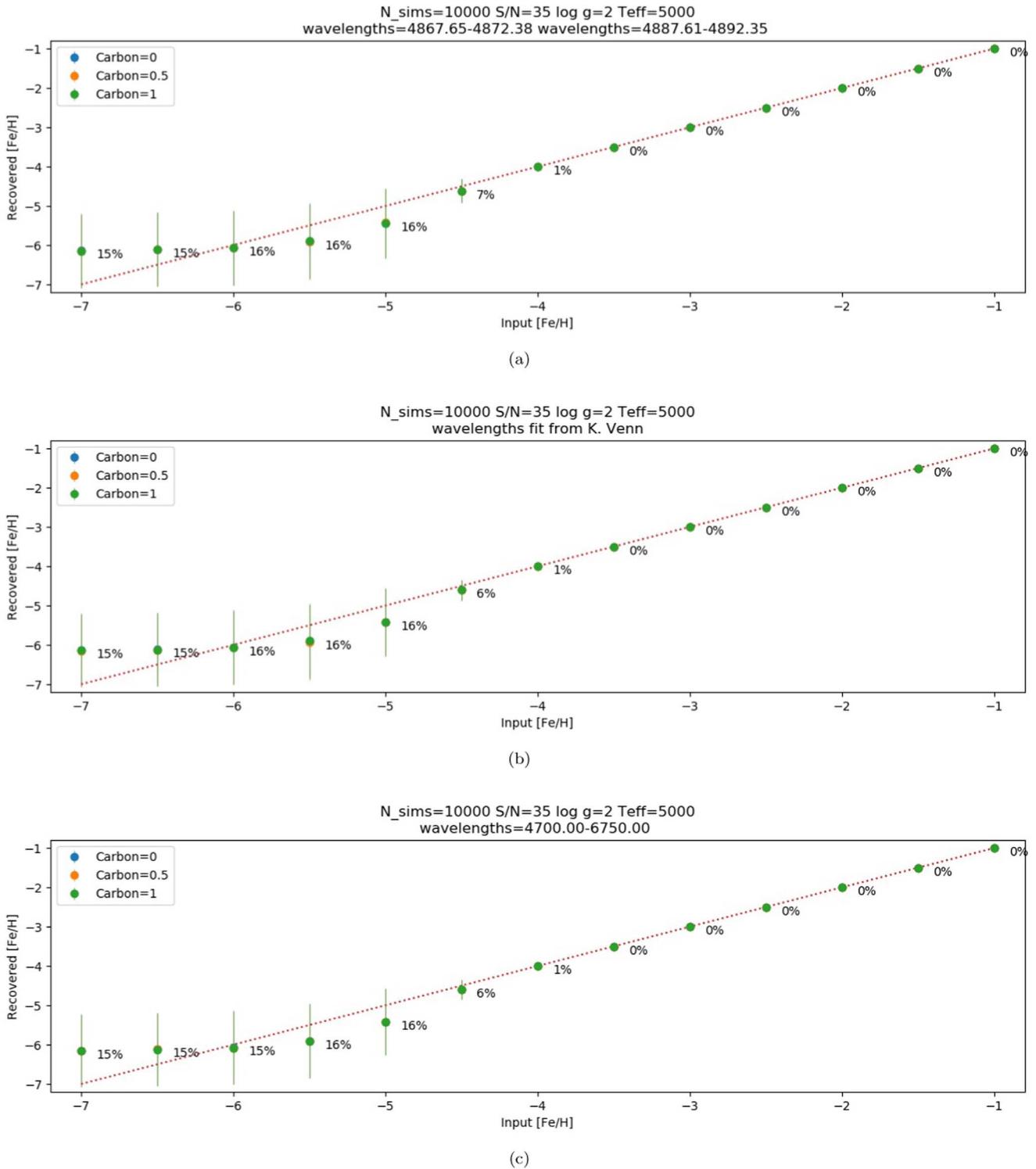

**Figure 18.** Same format as Figure 16, now with panels showing three different combinations of spectral lines: top is the Nordlander best features (Figure 17), middle is fit to 57 metal-sensitive features (K. Venn, private communication), and the bottom shows fitting results to the first three HERMES channels. The fit to three spectral channels yields a marginally better constraint at [Fe/H] ∼ -4.5 for cool giants with S/N = 35 than fits for [Fe/H] to the other spectra regions.

Figure 17 shows simulation runs where multiple spectral lines were simultaneously fit to illustrate improvement in metallicity constraints. While the second strongest set of lines improves the fitting to lower metallicities, the third set of lines does not influence the measured metallicity significantly for S/N = 35 spectra.

Figure 18 shows metallicity-sensitivity simulations for an additional line lists (K. Venn, private communication) with 57 features as well as full fits to the spectra over the three

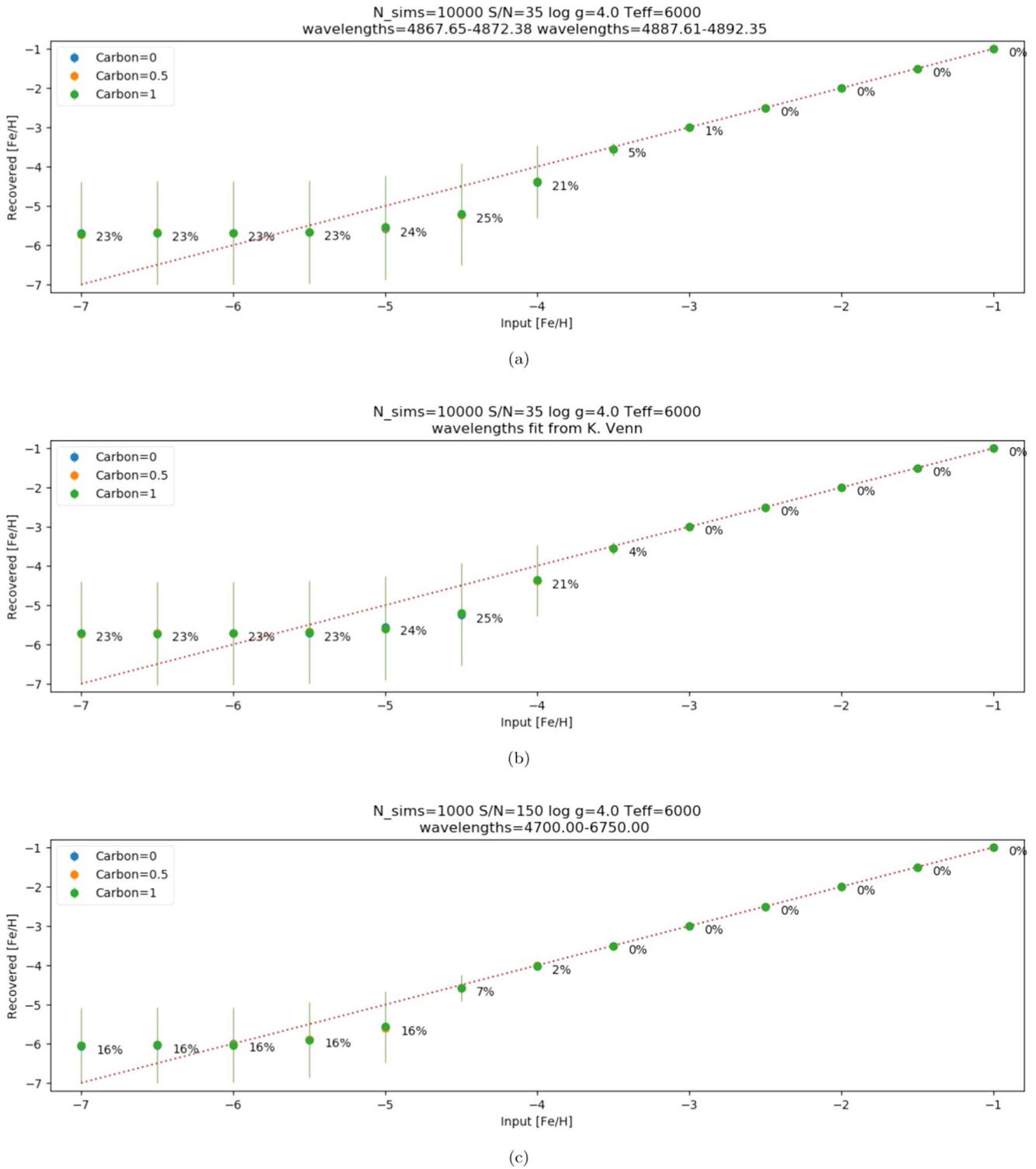

**Figure 19.** Same as Figure 18 now for hot ($T_{\rm eff} = 6000$) main-sequence ($\log g = 4$) stellar templates. Overall, the metallicity sensitivity decreases such that we may only expect to make measurements to [Fe/H] ∼ −3.5.

HERMES channels. These are compared to the best combination of two lines from Figure 17. Increased wavelength coverage appears to yield better metallicity constraints in the simulations, but only marginally so.

We show simulation results for hot main-sequence stars in Figure 19. At these temperatures, the metallicity sensitivity decreases, so that only metallicities of [Fe/H] ∼ −3.5 or higher are measurable with S/N = 35 GALAH spectra.

Finally, we show in Figure 20 how the metallicity sensitivity is expected to improve for higher S/N (S/N ∼150) data: [Fe/H] ∼ −5.0 and ∼−4 are expected for cool giants and hot main-sequence stars, respectively.

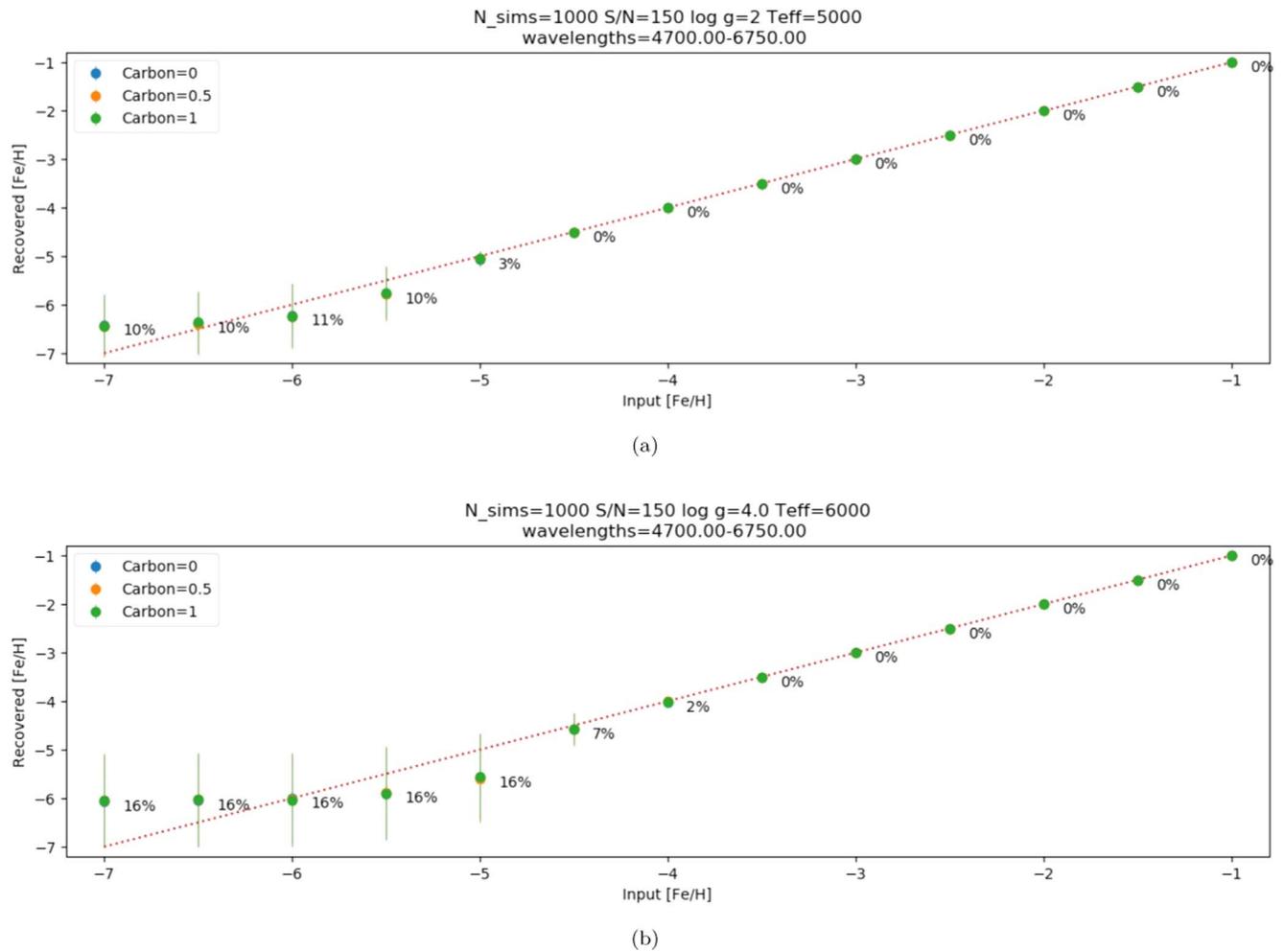

**Figure 20.** A few spectra in the GALAH sample reach S/N = 150. These simulations show how much better the low-metallicity constraints can be with higher S/N data and fits over the entire first three GALAH channels.

To conclude, we find the following:

1. Synthetic cool giant spectra with typical GALAH S/N = 35 over the GALAH spectral range are good (∼9%) at recovering metallicities as low as [Fe/H] ∼ −5.5.
2. At a certain metallicity the GALAH spectra are no longer sensitive to lower metallicities for S/N = 35 spectra.
3. With better S/N (∼100, or even ∼150), metallicities as low as [Fe/H] ∼ −4.5 can be recovered to ∼9%.
4. The metallicity sensitivity and fitting does not appear to be impacted by the level of carbon enhancement of the star within the wavelength coverage of the GALAH spectral channels.

## ORCID iDs

Arvind C. N. Hughes ● https://orcid.org/0000-0001-9294-3089
Lee R. Spitler ● https://orcid.org/0000-0001-5185-9876
Daniel B. Zucker ● https://orcid.org/0000-0003-1124-8477
Thomas Nordlander ● https://orcid.org/0000-0001-5344-8069
Jeffrey Simpson ● https://orcid.org/0000-0002-8165-2507
Gary S. Da Costa ● https://orcid.org/0000-0001-7019-649X
Yuan-Sen Ting ● https://orcid.org/0000-0001-5082-9536
Chengyuan Li ● https://orcid.org/0000-0002-3084-5157
Joss Bland-Hawthorn ● https://orcid.org/0000-0001-7516-4016
Sven Buder ● https://orcid.org/0000-0002-4031-8553
Andrew R. Casey ● https://orcid.org/0000-0003-0174-0564
Gayandhi M. De Silva ● https://orcid.org/0000-0001-7362-1682
Valentina D'Orazi ● https://orcid.org/0000-0002-2662-3762
Ken C. Freeman ● https://orcid.org/0000-0001-6280-1207
Michael R. Hayden ● https://orcid.org/0000-0001-7294-9766
Geraint F. Lewis ● https://orcid.org/0000-0003-3081-9319
Karin Lind ● https://orcid.org/0000-0002-8892-2573
Sarah L. Martell ● https://orcid.org/0000-0002-3430-4163
Katharine J. Schlesinger ● https://orcid.org/0000-0003-0110-0540
Sanjib Sharma ● https://orcid.org/0000-0002-0920-809X
Tomaž Zwitter ● https://orcid.org/0000-0002-2325-8763